\documentclass[11pt, a4paper]{article}	

\usepackage[english]{babel}
\usepackage[T1]{fontenc}
\usepackage[utf8]{inputenc}
\usepackage{mathtools} 		
\usepackage{amssymb}  		
\usepackage{upgreek}
\usepackage{graphicx} 	
\usepackage{xcolor} 
\definecolor{Blu}{cmyk}{1,0.6,0,0.2}	
\usepackage{booktabs}

\usepackage{authblk}
\usepackage{cite}

\usepackage{hyperref}
  \hypersetup{linktoc=all, breaklinks, hidelinks, colorlinks=true, linkcolor=Blu, citecolor=Blu, urlcolor=Blu,}
\usepackage{bookmark} 

\newcommand*{\iu}{\mathrm{i}\mkern1mu}  
\newcommand*{\e}{\mathrm{e}} 
\newcommand*{\ddelta}{\updelta}
\newcommand*{\mat}[1]{\begin{pmatrix} #1 \end{pmatrix}}
\newcommand*{\vb}[1]{\boldsymbol{#1}}
\newcommand*{\dd}[1]{\mathrm{d}#1}
\newcommand*{\N}{\mathbb{N}}
\newcommand*{\Z}{\mathbb{Z}}	
\newcommand*{\R}{\mathbb{R}}
\newcommand*{\C}{\mathbb{C}}
\newcommand*{\UU}{\mathrm{U}}	
\DeclareMathOperator{\sign}{sign}
\DeclareMathOperator{\sinc}{sinc}
\DeclareMathOperator{\tr}{tr}
\DeclareMathOperator{\arctanh}{arctanh}

\DeclareMathOperator{\spec}{spec}
\DeclareMathOperator{\id}{id}

\renewcommand\Im{\operatorname{Im}}

\title{Zak phase and bulk-boundary correspondence in a generalized Dirac--Kronig--Penney model}

\date{}

\author[$\hspace{0cm}$]{Giuliano Angelone$^{1,}$\footnote{Corresponding author:  \href{mailto:giulianoangelone@gmail.com}{giulianoangelone@gmail.com}}}
\author[$\hspace{0cm}$]{Domenico Monaco$^{1}$}
\author[$\hspace{0cm}$]{Gabriele Peluso$^{1}$}

\affil[1]{Dipartimento di Matematica, Sapienza Università di Roma\newline Piazzale Aldo Moro 5, 00185 Rome, Italy}

\begin{document}
\maketitle

\begin{abstract}
We investigate the topological properties of a generalized Dirac--Kronig--Penney model, a continuum one-dimensional model for a relativistic quantum chain. By tuning the coupling parameters this model can accommodate five Altland--Zirnbauer--Cartan symmetry classes, three of which (AIII, BDI and D) support non-trivial topological phases in dimension one.
We characterize analytically the spectral properties of the Hamiltonian in terms of a spectral function, and numerically compute the Zak phase to probe the bulk topological content of the insulating phases. Our findings reveal that, while the Zak phase is quantized in classes AIII and BDI, it exhibits non-quantized values in class D, challenging its traditional role as a topological marker in continuum settings.
We also discuss the bulk-boundary correspondence for a truncated version of the chain,  analyzing how the emergence of edge states depends on both the truncation position and the  boundary conditions. In classes AIII and BDI, we find that the Zak phase effectively detects edge states as a relative boundary topological index, although the correspondence is highly sensitive to the parameters characterizing the truncation.
\end{abstract}

\section{Introduction}
Topology has become a prominent paradigm in condensed matter physics, as topologically protected edge states see applications in many quantum technologies, ranging from quantum computers~\cite{Ki01} to ultra-efficient electronic devices with minimal heat generation, exploiting the dissipationless currents that flow at the boundary of a topological insulator~\cite{HaKa, Chen12}. From a theoretical viewpoint, the seminal works of Kitaev and of Ryu, Schnyder, Furusaki and Ludwig classified the ``periodic table'' of topological phases in insulators and superconductor~\cite{Kit09,RSFL10, ChTeSc16}, providing sets of topological labels depending on the dimension and on the Altland--Zirnbauer--Cartan (AZC) \emph{symmetry class}~\cite{AlZi97, HeHuZi}. Their findings are summarized in Tab.~\ref{table}, where we report only the column corresponding to dimension $D=1$ which is of interest for the present work.

\begin{table}[tp]
\centering
\begin{tabular}{c c c c c}
    \toprule
AZC class & $T$ & $C$ & $S$ & $D=1$ \\
    \midrule
A & 0&0&0& 0\\
AIII& 0&0&1 &$\mathbb{Z}$ \\
AI    &1 & 0&  0 & 0 \\
BDI   &1 & 1&  1 & $\mathbb{Z}$ \\
D     &0 & 1&  0 & $\mathbb{Z}_2$ \\
DIII  &$-1$& 1&  1 & $\mathbb{Z}_2$ \\
AII   &$-1$& 0&  0 & 0 \\
CII   &$-1$&$-1$& 1  &$\mathbb{Z}$ \\
C     &0 &$-1$&  0 &0 \\
CI    &1 &$-1$&  1 &0 \\
    \bottomrule
\end{tabular}
\caption{
Classification of topological insulators and superconductors in dimension $D=1$.}
\label{table}
\end{table}

Table~\ref{table} should be read as follows. Consider a quantum chain, modeled by a Hamiltonian $H$  on a Hilbert space $\mathfrak{H}$. Finite symmetries of the system come in the following types.
\begin{itemize}
    \item \emph{Time-reversal symmetry} $T$: an anti-unitary operator such that $T^2=\pm \id_{\mathfrak{H}}$ and $HT=TH$.
    \item \emph{Charge-conjugation symmetry} $C$: an anti-unitary operator such that $C^2=\pm\id_{\mathfrak{H}}$ and $CH=-HC$.
    \item \emph{Chiral symmetry} $S$: a unitary operator such that $S^2=\id_{\mathfrak{H}}$ and $SH=-HS$.
\end{itemize}
In the table, when a symmetry is absent (i.e.\ broken), the corresponding column has an entry marked with a $0$; a $1$ or $-1$ denotes instead the presence of the  corresponding symmetry, as well as the value of the square of the corresponding operator. 
In each of the resulting ten symmetry classes, the topological index which can be associated to different Hamiltonians takes values in the groups $\{0\}$, $\Z_2$ or $\Z$, and provides invariants under continuous deformations of the Hamiltonian which respect the symmetry class and do not close the relevant spectral gap characterizing the insulating phase.
Different values of the index label the different topological quantum phases that can exist within a given symmetry class.
Such bulk topological indices, moreover, are expected to predict edge properties of a truncated version of the system, via the celebrated \emph{bulk-boundary correspondence} (BBC).

The overall picture presented above is actually more nuanced. Especially in presence of charge-conjugation and/or chiral symmetries, the topological indices do not give absolute labels to each quantum phase; rather, they acquire a \emph{relative} interpretation, and they should be attached to \emph{differences} between phases~\cite{Kit09, Thiang15}. 
Moreover, although the table predicts which set of indices label the different topological phases in a given symmetry class, they do not provide explicit information on any given model: one needs to find appropriate \emph{topological markers} which allow to compute the specific value of the system's indices.
Finally, while the notion of ``continuous deformation'' (i.e.\ homotopy) has a clear meaning in lattice tight-binding models, which rely on a finite-band truncation, in continuum models the presence of an infinite-dimensional (fiber) Hilbert space challenges the definition of topological invariants~\cite{Pe26}, and the survival of topological properties in the transition between continuum and discrete models has been questioned~\cite{ShWe22}. 

Although we will not tackle these issues in full generality, in this paper we propose a continuum model for a relativistic quantum chain as a playground to explore various topological phases and their properties, allowing us to probe the latter in an infinite-dimensional setting.
Specifically, the model that we introduce in Section~\ref{sec:model} and Appendix~\ref{app:pointint} is a \emph{generalized Dirac--Kronig--Penney} (gDKP) Hamiltonian $H_U$, consisting of a one-dimensional massive Dirac operator perturbed by a $\Z$-periodic array of point interactions which are characterized by a $\UU(2)$-valued coupling $U$.
Depending on the choice of such coupling matrix, all three symmetries (time-reversal, charge-conjugation, and chiral)  squaring to the identity can be accommodated, so that by tuning the ``strength'' of the point interaction one can explore classes A, AI, AIII, D, and BDI according to the AZC labels in Tab.~\ref{table}; the latter three classes are expected to host non-trivial topological phases according to the periodic table. 
The most commonly adopted topological marker for isolated spectral bands in one-dimensional systems is the \emph{Zak phase}~\cite{Zak89}, namely the Berry phase picked up by a Bloch function when it is ``transported'' along the Brillouin zone (see Appendix~\ref{App:BerryZak}). 
Non-trivial topological phases are usually characterized by a quantization modulo~$2\pi$ of the (relative) Zak phase. Whenever the quantum chain modeled by $H_U$ is insulating, we compute numerically the Zak phase of the energy bands close to zero, to probe its topological content (e.g.\ in term of its quantization). This analysis is carried out in Section~\ref{sec:bulk}. Section~\ref{sec:BulkBoundary} is instead devoted to studying a truncated version of the quantum chain and to establishing the validity or the violation of the BBC, which is formulated in terms of a boundary topological index associated to the number of edge states with energies in the bulk gaps, see Eq.~\eqref{eq:NbNa}. In particular, we extensively study how  such edge states are affected by the possible ways of performing the truncation, depending on its position  within the unit cell and on the additional boundary condition at the chain edge.

The gDKP model studied in this paper thus offers a rich display of topological phenomena, stimulating further studies to get a thorough understanding of one-dimensional topological phases of matter and their interplay with point interactions.

\paragraph{Acknowledgments.} 
The authors gratefully acknowledge financial support from Ministero dell’Università e della Ricerca (MUR, Italian Ministry of University and Research) and Next Generation EU within PRIN 2022AKRC5P ``Interacting Quantum Systems: Topological Phenomena and Effective Theories'' and within PNRR–MUR Project no.~PE0000023-NQSTI. The work of D.\ M.\ and G.\ P.\ was also supported by Sapienza Università di Roma within Progetto di Ricerca di Ateneo 2023 and 2024.

\section{Generalized Dirac--Kronig--Penney model} \label{sec:model}
\begin{figure}[tp]
\centering
\includegraphics{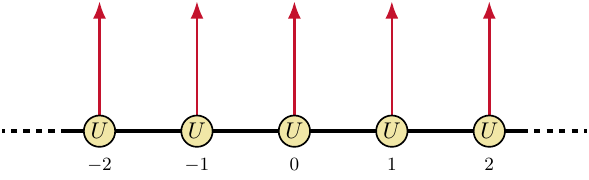}
\caption{Schematic representation of the gDKP model. At each lattice point $n\in \Z$ there is a point interaction realizing the $\UU(2)$ boundary condition $\Psi_-(n)=U\Psi_+(n)$.}
\label{fig:comb}
\end{figure}

We consider a relativistic (spin-$\tfrac{1}{2}$) quantum particle of mass $m>0$, moving in a one-dimensional periodic array of point interactions which are placed at the lattice points $n\in \Z$. We restrict our attention to $\C^2$-valued \emph{spinors}
\begin{equation*}
    \Psi(x)=\mat{\phi(x)\\ \chi(x)}\in L^2(\R;\C^2),
\end{equation*}
although one could also consider $\C^4$-valued spinors~\cite{McFa04, AkBe08, CaNeGui09, diracred}. The most general self-adjoint Hamiltonian describing this system, in the Dirac representation~\cite{isodirac} and adopting natural units $\hbar=c=1$, is the following gDKP model~\cite{AvGr76, GeSe87, Yos09} (or \emph{relativistic Dirac comb}): for any $2\times 2$ unitary matrix $U\in \UU(2)$ let
\begin{align}\label{eq:HU}
    H_U=-\iu\sigma_x\frac{\dd}{\dd x}+ m\sigma_z
\end{align}
be defined in
\begin{equation}\label{eq:DHU}
    \mathfrak{D}(H_U)=\bigl\{\Psi \in H^1(\R\setminus\Z; \C^2) : \Psi_{-}(n)=U\Psi_{+}(n)\,\forall\, n\in\Z\bigr\},
\end{equation}
where $H^1(\Omega)$ is the space of functions with square-integrable weak derivative (that is, the first Sobolev space) with support in $\Omega$ while
\begin{equation}\label{eq:Psipm}
    \Psi_{\pm}(n)= \frac{1}{\sqrt{2}}\mat{ \phi(n^-) \pm \chi(n^-) \\[4pt] \phi(n^+) \mp \chi(n^+) },
\end{equation}
are two vectors containing the boundary data of $\Psi(x)$ at the lattice points. The domain~\eqref{eq:DHU} enforces the following (spin-dependent) \emph{coupling conditions}
\begin{equation*}
\mat{ \phi(n^-) - \chi(n^-) \\[2pt] \phi(n^+) + \chi(n^+) }=U\mat{ \phi(n^-) + \chi(n^-) \\[2pt] \phi(n^+) - \chi(n^+) },
\end{equation*}
which are the same at all the lattice points $n\in \Z$, as sketched in Fig.~\ref{fig:comb}. 
From a physical perspective, the point interactions act as zero-range barriers whose microscopic structure is entirely captured by the matrix $U$, and indeed the above coupling conditions admit a clear physical interpretation in terms of local scattering matrices.

We stress that this $\UU(2)$ family of Hamiltonians naturally emerges as consequence of von Neumann’s extension theory for one-dimensional Dirac operators with deficiency indices $(2,2)$. Different choices of the coupling matrix $U\in\UU(2)$ correspond to different physical systems, and each coupling condition realize a particular kind of point interaction. The Hamiltonian $H_U$ is indeed strictly related with the more familiar (singular) Dirac operator with a periodic array of Dirac $\ddelta$-potentials, namely
\begin{align}\label{eq:Hg}
    \tilde{H}_{\vb{g}}=-\iu\sigma_x\frac{\dd}{\dd x}+m\sigma_z +V_{\vb{g}}\sum_{n\in \Z}\ddelta(x-n)
\end{align}
where $V_{\vb{g}}$ is the Hermitian matrix 
\begin{align}
V_{\vb{g}}=\mat{g_0+g_3 & g_1-\iu g_2 \\  g_1+\iu g_2 &  g_0- g_3}
\end{align}
containing the coupling parameters $\vb{g}=(g_0,g_1,g_2,g_3)\in\R^4$. The precise definition of $\tilde{H}_{\vb{g}}$, which is the most general \emph{singular perturbation} of the free Dirac operator with periodic point interactions, requires the generalization of distribution theory to discontinuous functions~\cite{AlbKur97, Kur96, HeTu22, CLMT23}, and it is not relevant for our scopes. Here we just mention that if $1\notin \spec(U)$, i.e.\ if the matrix $U$ has no eigenvalues equal to $1$, one can find a one-to-one \emph{Kurasov mapping} $U\mapsto\vb{g}(U)$ such that the action of the two operators coincide, that is
\begin{equation*}
    H_{U}=\tilde{H}_{\vb{g}(U)},
\end{equation*}
see Appendix~\ref{app:pointint} for further details. On the other hand, if $1\in \spec(U)$, the correspondence fails to be bijective, and there are different operators $H_U$ that formally correspond to a singular Dirac operator $\tilde{H}_{\vb{g}(U)}$ with (possibly) infinite couplings $\vb{g}$, that is with $\vb{g}(U)\in(\R\cup\{\infty\})^4$. Similar relations hold also in the non-relativistic case~\cite{Kur96, Lan15, isocomb}.

The analysis of one-dimensional Dirac operators with point interactions has a long history, both in the physical and mathematical literature~\cite{Lap83, CaKiNo87, BeDa94, AlbKur97, Hug99, SuJa04,  CaMaPo13, GuMuPi19, BHCS24}, although in our opinion the most general periodic model leading to the $\UU(2)$ family in Eq.~\eqref{eq:HU} has received only a mild attention~\cite{AlDeV00, PaRi14,HeTu22,  BoLuMa24, isodirac}. Our interest in this system lies in the fact that it provides a non-trivial instance of continuum models describing one-dimensional \emph{topological insulators}. More precisely, the Hamiltonian $H_U$ can belong to different AZC symmetries classes~\cite{AlZi97,HeHuZi} associated to the classification of topological matter~\cite{Kit09,RSFL10}, depending on the particular choice of the coupling matrix $U\in\UU(2)$.
We mention that non-trivial topological properties have already been observed in some variations of the Kronig--Penney model~\cite{RaKe05,  HoWeRa11, RBRB19, SmKoPr21, NZHFB25}. However, non-relativistic models can realize only a limited range of topological symmetry classes, since Schrödinger operators are usually bounded from below, and thus, they cannot be endowed with charge-conjugation  or chiral symmetry. On the other hand, relativistic models described by Dirac operators are typically not semi-bounded, and can thus explore a wider range of symmetry classes.

Our model, in particular, contains representatives from all the classes whose symmetry operators square to the identity, namely the classes A, AI, AIII, D and BDI.\footnote{We expect that by extending the model to four-component spinors one could also consider symmetry operators squaring  to minus the identity.} We discuss the relation between coupling conditions and symmetry classes in the next section: we anticipate here that, by ``tuning'' the coupling  $U$, we can transition between different topological phases in the same symmetry class, and in certain cases also from class to class. In Section~\ref{sec:coupcond} we then highlight some physical properties of the coupling conditions, also providing some relevant examples.

\subsection{Symmetry classes}\label{sec:symmetryclass}
Let $K\colon \Psi(x)\mapsto \overline{\Psi(x)}$ be the anti-unitary operator of complex conjugation on $L^2(\R;\C^2)$, and let
\begin{align*}
T=\iu\sigma_z K, && C=\sigma_x K, && S=TC=-\sigma_y,
\end{align*}
where $(\sigma_x,\sigma_y,\sigma_z)$ denote the Pauli matrices. Notice that $T$ and $C$ are anti-unitary, while $S$ is unitary. These operators satisfy the identity
\begin{align*}
T^2=C^2=S^2=\id_{L^2(\R;\C^2)},
\end{align*}
as well as the following (anti-)commutation relations
\begin{align*}
TH=HT, && CH=-CH, && SH=-SH,
\end{align*}
with the free Dirac operator given by
\begin{align}\label{eq:Hfree}
H=-\iu\sigma_x\frac{\dd}{\dd x}+ m\sigma_z\,,&& \mathfrak{D}(H)=H^1(\R; \C^2).
\end{align}
Therefore, $T, C$ and $S$ represent respectively the operators of time-reversal, charge-conjugation and chiral symmetry. Let us remark that even though the free Hamiltonian $H$ is invariant with respect to all of the above symmetries, the point interactions contained in the gDKP model $H_U$ can break any of them in a non-trivial way.
In order to discuss the effect of each symmetry on $H_U$, let us recall that  a generic $\UU(2)$ matrix can be parametrized by
\begin{align}\label{eq:Uparam}
    U=\e^{\iu\eta}\mat{m_0+\iu m_3 & m_2+\iu m_1 \\ -m_2+\iu m_1 & m_0-\iu m_3}
\end{align}
where $\eta\in[0,\pi)$ while the real parameters $\{m_i\}_{0\le i\le 3}$ are subjected to the $\mathbb{S}^3$ constraint $m_0^2+m_1^2+m_2^2+m_3^2=1$.  

Under the time-reversal operator $T$ the boundary vectors in Eq.~\eqref{eq:Psipm} transform as $\Psi_{\pm}(n) \mapsto \iu\overline{\Psi}_{\mp}(n)$. This means that if $\Psi\in\mathfrak{D}(H_U)$ satisfies the coupling conditions $\Psi_{-}(n)=U\Psi_{+}(n)$, then $\Phi\coloneq T\Psi\in T\mathfrak{D}(H_U)$ satisfies the new coupling conditions $\Phi_{-}(n)=U^\intercal\Phi_{+}(n)$, where $U^\intercal$ is the transpose of $U$. This in turn implies the anti-unitary equivalence 
\begin{align*}
	TH_UT^{-1} =H_{U^\intercal}
\end{align*}
between the two generally different Hamiltonians $H_U$ and $H_{U^\intercal}$. We can conclude that the gDKP model will be be invariant with respect to $T$ if and only if $U=U^\intercal$, which in terms of the parameters introduced in Eq.~\eqref{eq:Uparam} reads
\begin{align}\label{eq:Tcond}
	m_2=0.
\end{align}
Analogously, under the charge-conjugation operator $C$ the boundary vectors transform as $\Psi_{\pm}(n) \mapsto \pm \sigma_z\overline{\Psi}_{\pm}(n)$, realizing the anti-unitary equivalence 
\begin{align*}
	CH_{U} C^{-1} =- H_{-\sigma_z \overline{U}\sigma_z }.
\end{align*}
Thus, the Hamiltonian $H_U$ will be symmetric with respect to $C$ if and only if 
\begin{align}\label{eq:Ccond}
     \eta=0,\,m_0=m_1=0 \qquad \text{or} \qquad     \eta=\frac{\pi}{2},\,m_2=m_3=0 .
\end{align}
The chiral operator $S$, in turn, acts as $\Psi_{\pm}(n) \mapsto \iu \sigma_z\Psi_{\mp}(n)$ realizing the unitary equivalence 
\begin{align*}
	SH_{U} S^{-1} =- H_{-\sigma_z U^\dagger\sigma_z}\,.
\end{align*}
We conclude that $H_U$ is symmetric with respect to $S$ if and only if
\begin{align}\label{eq:Scond}
      \eta=0,\,m_0=m_1=m_2=0 \qquad \text{or} \qquad     \eta=\frac{\pi}{2},\,m_3=0.
\end{align}

As the reader can easily verify, the parameter space of our model is large enough to accommodate any of the symmetry classes A, AI, AIII, D and BDI. In the rest of this paper we will focus only on the classes exhibiting  non-trivial topological features in dimension one, namely class D (with charge-conjugation  symmetry) having a $\Z_2$ index, and classes BDI (with charge-conjugation and chiral symmetry) and AIII (with chiral symmetry), both having  a $\Z$ index. Exploiting Eqs.~\eqref{eq:Tcond}--{\eqref{eq:Scond}, we say that the Hamiltonian $H_U$ is in class D if
\begin{align}\label{eq:UD}
 U\in \mathcal{U}_{\text{D}}=\bigl\{ U_C(\theta):\theta\in [-\pi,\pi)\bigr\}
\end{align}
where we set
\begin{align}\label{eq:UC}
 U_C(\theta)=\mat{\iu\cos(\theta) & \sin(\theta)  \\ -\sin(\theta) & -\iu\cos(\theta)}.
\end{align}
Let us remark that for $U_{C}(-\pi)=-\iu\sigma_z$ and $U_C(0)=\iu\sigma_z$ the Hamiltonian is symmetric with respect to both charge-conjugation and chiral symmetry, and thus strictly speaking  $H_{U_{C}(-\pi)}$ and $H_{U_{C}(0)}$ are in class BDI. Nevertheless, we can connect any two Hamiltonians associated to the set of coupling conditions $\mathcal{U}_{\text{D}}$ by means of a continuous transformation that does not break the charge-conjugation symmetry. Excluding the isolated matrices $U=\pm\iu\sigma_z$, $H_U$ is in the BDI class if
\begin{align}\label{eq:UBDI}
 U\in \mathcal{U}_{\text{BDI}}= \bigl\{ U_{CS}(\theta):\theta\in [-\pi,\pi)\bigr\} 
\end{align}
where
\begin{align}\label{eq:UCS}
U_{CS}(\theta)=\mat{\iu\cos(\theta) & -\sin(\theta)  \\ -\sin(\theta) & \iu\cos(\theta)}.
\end{align}
Finally, we say that $H_U$ is in the AIII class  if
\begin{align}\label{eq:UAIII}
 U\in \mathcal{U}_{\text{AIII}}= \bigl\{ U_{S}(\theta,m_2): (\theta,m_2)\in [-\pi,\pi)\times [-1,1]\bigr\}
\end{align}
where
\begin{align}\label{eq:US}
  U_{S}(\theta,m_2)=\mat{\iu\sqrt{1-m_2^2}\cos(\theta) & \iu m_2-\sqrt{1-m_2^2}\sin(\theta) \\ -\iu m_2- \sqrt{1-m_2^2}\sin(\theta) &\iu \sqrt{1-m_2^2}\cos(\theta)}.
\end{align}
As before, we included in $\mathcal{U}_{\text{AIII}}$ also the matrices $U_S(\theta,0)=U_{CS}(\theta)$ that belong to $\mathcal{U}_{\text{BDI}}$: in this way we look at the set $\mathcal{U}_{\text{BDI}}$ as continuously embedded in $\mathcal{U}_{\text{AIII}}$ by fixing the value of the parameter $m_2=0$. We stress that this embedding preserves chiral symmetry, but explicitly breaks both charge-conjugation and time-reversal symmetry.

\subsection{Permeable and impermeable coupling conditions}\label{sec:coupcond}
The probability current density of a spinor $\Psi(x)$ is given, for the Dirac operator in Eq.~\eqref{eq:HU}, by the expression
\begin{equation*}
j_\Psi(x)=\iu \Psi^\dagger(x)\sigma_x\Psi(x)=\iu \bigl(\overline{\phi(x)}\chi(x)+\overline{\chi(x)}\phi(x)\bigr)
\end{equation*}
where $\Psi^\dagger(x)$ is the adjoint vector of $\Psi(x)$. In accordance with the unitarity of the evolution and the conservation of probability enforced by the self-adjointness of the Hamiltonian, the coupling conditions $\Psi_-(n)=U\Psi_+(n)$ ensure that $j_\Psi(n^+)=j_\Psi(n^-)$ for any $n\in \Z$. In particular, depending on the behavior of $j_\Psi(x)$ at the lattice points, coupling conditions fall into two classes with different physical properties: impermeable or permeable. 

A coupling condition is \emph{impermeable} (or confining) if
\begin{equation*}
j_\Psi(n^\pm)=0.
\end{equation*} 
Impermeable conditions do not allow the propagation of probability across any of the lattice points, effectively decoupling the system into an infinite collection of equal segments. The most general impermeable condition is obtained by letting
\begin{equation}\label{eq:permcond}
m_1=m_2=0
\end{equation}
in Eq.~\eqref{eq:Uparam}, leading to the diagonal coupling matrices
\begin{equation*}
U_\text{ch}(\alpha_-,\alpha_+)=\mat{\e^{\iu\alpha_-} & 0 \\ 0 & \e^{\iu\alpha_+}}
\end{equation*}
where $\alpha_\pm\in[-\pi,\pi)$. The corresponding boundary conditions
\begin{align*}
\sin(\tfrac{\alpha_\pm}{2})\phi(n^\pm)=\pm\iu \cos(\tfrac{\alpha_\pm}{2})\chi(n^\pm).  
\end{align*}
are known as \emph{chiral conditions}~\cite{JaMa89,HoTo96,RoYa21,isodirac}, and can also be rewritten as
\begin{align*}
\bigl(I-\sigma_z\e^{\pm\iu \alpha_\pm \sigma_x} \bigr)\Psi(n^\pm)=0.
\end{align*}

Vice versa, a coupling condition is \emph{permeable} (or connected) if 
\begin{equation*}
j_\Psi(n^\pm)\neq 0,
\end{equation*} 
allowing the propagation of probability across the lattice points. A well-known example is given by the \emph{pseudo-periodic conditions}
\begin{equation*}
\Psi(n^+)=\e^{\iu\alpha}\Psi(n^-)
\end{equation*}
obtained from the anti-diagonal matrices
\begin{equation*}
U_\text{pp}(\alpha)=\mat{0 & \e^{-\iu\alpha} \\ \e^{\iu\alpha} & 0}
\end{equation*}
where $\alpha\in[-\pi,\pi)$. Pseudo-periodic conditions are associated to a singular gauge field. By exploiting the Kurasov mapping~\eqref{eq:gU} derived in Appendix~\ref{app:pointint}, we find indeed that $H_{U_\text{pp}(\alpha)}$ has the same action of
\begin{equation*}
	\tilde{H}_{\vb{g}(U_{\text{pp}}(\alpha))}=-\iu\sigma_x\biggl(\frac{\dd}{\dd x}
	-2\iu\tan\Bigl(\frac{\alpha}{2}\Bigr) \sum_{n\in\Z}\ddelta(x-n)\biggr)+m\sigma_z.
\end{equation*}
Notice that pseudo-periodic conditions can be \emph{gauged away}~\cite{CLMT23,isocomb}, i.e.\ 
the Hamiltonian $H_{U_{\text{pp}}(\alpha)}$ is unitarily equivalent to the free Dirac operator in Eq.~\eqref{eq:Hfree}  via the singular gauge transformation
\begin{equation*}
	\Psi(x)\mapsto\e^{-\iu \alpha\sum_{n\in \Z}  \Uptheta(x-n)-\Uptheta(-n)}\Psi(x),
\end{equation*}
where $\Uptheta(x)$ is the Heaviside step function (with $\Uptheta(0)=1$). In particular, this implies the isospectrality relation
\begin{align*}
\spec\bigl(H_{U_{\text{pp}}(\alpha)}\bigr)=\spec(H)=(-\infty,m]\cup[m,+\infty)
\end{align*}
for any $\alpha\in[-\pi,\pi)$.

\section{Bulk analysis}\label{sec:bulk}
In this section we study both the spectral and topological properties of the gDKP model $H_U$. Outlining the exact band structure of $H_U$ is indeed essential to determine the Zak phase associated to each energy band, and to discuss the topological properties encoded in the latter. As we will show in Section~\ref{sec:spectral}, by exploiting the Bloch decomposition the spectral problem of $H_U$ is exactly solvable in terms of a spectral function. Then in Section~\ref{sec:topological} we analyze the behavior of the Zak phase for the classes D, BDI and AIII, determining whether they admit any topological phase transition.

\subsection{Spectral properties}\label{sec:spectral}
We can take advantage of the (discrete) translation invariance of $H_U$ by exploiting the Bloch–Floquet transform $\mathcal{U}_\text{BF}$~\cite{ReSi78,Ku16}, which is defined for any $\Psi(x)$ in the Schwartz space $\mathcal{S}(\R;  \C^2)$ by
\begin{align*}
(\mathcal{U}_\text{BF}\Psi)(k,x)=\sum_{n\in\Z} \e^{-\iu k n}\Psi(x-n).
\end{align*}
and extended uniquely to a unitary operator
\begin{align*}
\mathcal{U}_{\text{BF}}\colon L^2(\R;\C^2)\to \int_{[-\pi,\pi)}^\oplus L^2((-\tfrac{1}{2},\tfrac{1}{2});\C^2)\,\dd{k} 
\end{align*}
where the measure $\dd{k}$ is suitably normalized. Let us recall that when extended as a function of $L^2_{\text{loc}}(\R;\C^2)\otimes L^2_{\text{loc}}(\R;\C^2)$,  the Bloch transformed spinor $\Phi(k,x)\coloneq (\mathcal{U}_{\text{BF}}\Psi)(k,x)$ is pseudo-periodic in real space and periodic in reciprocal space, that is
\begin{align*}
\Phi(k,x+n)=\e^{\iu k n}\Phi(k,x),&&\Phi(k+2\pi n,x)=\Phi(k,x),
\end{align*}
for any $n\in\Z$ and for all $k,x\in\R$. In the Bloch--Floquet representation, $H_U$ admits the following fiber decomposition
\begin{align}\label{eq:hU}
    \mathcal{U}_\text{BF} H_U \mathcal{U}_\text{BF}^\dagger =\int_{[-\pi,\pi)}^\oplus h_U(k)\,\dd{k},&&h_U(k)=-\iu\sigma_x\frac{\dd}{\dd x}+ m\sigma_z,
\end{align}
where the domain of  $h_U(k)$ can be written as
\begin{equation*}
\begin{aligned}
     \mathfrak{D}(h_U(k))=\{\Psi \in H^1((-\tfrac{1}{2},\tfrac{1}{2})\setminus\{0\};\C^2) : {} & 
     \Psi(\tfrac{1}{2})=\e^{\iu k}\Psi(-\tfrac{1}{2}),\\&\, \Psi_{-}(0)=U\Psi_{+}(0)
     \}  .
\end{aligned}
\end{equation*}
As depicted in Fig.~\ref{fig:ring}, the fiber Hamiltonian $h_U(k)$ describes a free relativistic particle in a ring with two point interactions placed at antipodal points, one coming from the Bloch--Floquet pseudo-periodic condition $\Psi(\tfrac{1}{2})=\e^{\iu k}\Psi(-\tfrac{1}{2})$, the other from the $\UU(2)$ coupling condition $\Psi_{-}(0)=U\Psi_{+}(0)$.

\begin{figure}[tp]
\centering
\includegraphics{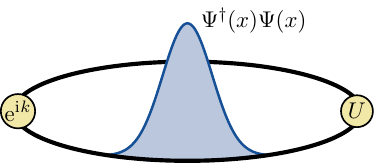}
\caption{The physical system described by the fiber Hamiltonian $h_U(k)$ is a ring with two point interactions at antipodal points. Here, the left and right point interactions implement respectively the Bloch--Floquet condition $\Psi(\tfrac{1}{2})=\e^{\iu k}\Psi(-\tfrac{1}{2})$ and  the $\UU(2)$ coupling condition $\Psi_-(0)=U\Psi_+(0)$.}
\label{fig:ring}
\end{figure}

\subsubsection{Spectral function}
The operator $h_U(k)$ is not semi-bounded, and its spectrum consists of countably many eigenvalues accumulating to both $+\infty$ and $-\infty$. We will denote these eigenvalues by $\epsilon_{U,n}(k)$ where $n\in\Z$ or $n\in \Z^\ast \coloneq \Z \setminus \{0\}$, depending respectively on the presence or absence of the zero eigenvalue, and $\epsilon_{U,n}(k)\le \epsilon_{U,n+1}(k)$. From standard results~\cite{ReSi78} we  know that the spectrum of $H_U$ consists of infinitely many \emph{energy bands}, each one given by the closure of the set $\{\epsilon_{U,n}(k):k\in[-\pi,\pi)\}$, so that solving the spectral problem of $H_U$ is equivalent to solving the eigenvalue problem of $h_U(k)$ for each $k$ in the Brillouin zone $[-\pi,\pi)$. Notice that each eigenvalue of $h_U(k)$ can be at most doubly degenerate, and (unless the spectrum of $H_U$ is pure point) if $\epsilon_{U,n}(k)$ happens to be degenerate the \emph{energy gap} between the $n$-th and the $(n+1)$-th (or the $(n-1)$-th) energy band closes.

The spectrum of $h_U(k)$ can be analytically characterized in terms of the real zeros of a \emph{spectral function} $F_{U,k}(\epsilon)$, that is 
\begin{equation*}
\spec(h_U(k))=\{\epsilon\in\R:F_{U,k}(\epsilon)=0\}.
\end{equation*}
In Appendix~\ref{app:spectral} we derive the explicit expression
\begin{equation}\label{eq:specfun}
    F_{U,k}(\epsilon)=m_1 \cos(k) + m_2\sin(k)+\cos(q)\sin(\eta)+\sinc(q)(\epsilon\cos(\eta)-mm_0)
\end{equation}
written in terms of the parametrization~\eqref{eq:Uparam} of  $U\in\UU(2)$, where
\begin{equation}\label{eq:wavenumber}
    q=q(\epsilon)=\e^{\iu \arg(\epsilon^2-m^2)/2}\sqrt{|\epsilon^2-m^2|}
\end{equation}
is a non-negative real number outside of the \emph{mass gap}, that is for $|\epsilon| \ge m$, and a purely imaginary number for $\epsilon\in(-m,m)$. 
The spectral condition $F_U(k,\epsilon)=0$ represents the relativistic counterpart of the Kronig--Penney  relation~\cite{FaGlSt73, Dom87}, generalized to the full $\UU(2)$ set of self-adjoint point interactions. 
By inspecting Eq.~\eqref{eq:specfun} we conclude that the spectrum of $H_U$ is either absolutely continuous or pure point if, respectively, $U$ is a permeable or impermeable coupling condition. In the second case, indeed, the constraint $m_1=m_2=0$ derived in Eq.~\eqref{eq:permcond} implies that all the energy bands are \emph{flat}, and each $\epsilon_{U,n}\in \spec(h_U(k))$ is an eigenvalue of $H_U$ with infinite degeneracy.
\subsubsection{Eigenspinors} \label{sec:Eigenspinors}
If $\epsilon=\epsilon_{U,n}(k)$ is a simple eigenvalue of $h_U(k)$ with $\epsilon\neq\pm m$, the corresponding eigenspinor is given by
\begin{equation}\label{eq:eigenspinor}
    \Psi_{U,n}(k,x)=\begin{cases}
    \xi_{U,n}^+
        \e^{\iu q x}
        +\xi_{U,n}^- \e^{-\iu q x},& x<0\\[2pt] 
        \xi_{U,n}^+ \e^{\iu [q(x-1)+k]}
        +\xi_{U,n}^- \e^{-\iu [q(x-1)-k]},& x>0
    \end{cases}
\end{equation}
where we introduced the vectors
\begin{align}
    \xi_{U,n}^\pm=c_{U,n}^\pm  \mat{1 \\ \pm \frac{q}{\epsilon+m}}
\end{align}
and the coefficients
\begin{align}
     c_{U,n}^\pm =\pm \bigl(1\pm \tfrac{q}{\epsilon+m}\bigr)(1-\e^{\iu (k+\eta\pm q)} (\iu m_1+m_2)) \mp \bigl(1\mp\tfrac{q}{\epsilon+m}\bigr)\e^{\iu \eta}(m_0+\iu m_3).
\end{align}
Analogous expressions can be derived also for $\epsilon=\pm m$, see Appendix~\ref{app:spectral} and~\cite{isodirac} for further details.

 \subsubsection{Zero eigenvalues}\label{sec:zeromodes}
The fiber Hamiltonian $h_U(k)$ has a zero eigenvalue if and only if
 \begin{align*}
     F_{U,k}(0)
     =m_1 \cos(k)+m_2\sin(k)+\cosh(m)\sin(\eta)-m_0 \sinh(m)
     =0.
 \end{align*}
For impermeable conditions the above equation reduces to
\begin{equation*}
     \sin(\eta)=m_0\tanh(m).
\end{equation*}
For permeable conditions, by introducing the auxiliary function
\begin{align}
    G(m,\eta,m_0)&=\cosh(m)\sin(\eta)-m_0 \sinh(m)
\end{align}
we can distinguish between three cases. The equation $F_{U,k}(0)=0$ has:
\begin{itemize}
\item no solutions if $|G(m,\eta,m_0)|>\sqrt{ m_1^2+m_2^2 }$;
\item a unique solution if $G(m,\eta,m_0)=\pm\sqrt{ m_1^2+m_2^2 }$, that is obtained for
\begin{equation*}
	k_\pm= 
	\begin{cases}
	\bigl(\pi-\arctan\bigl(\tfrac{m_1}{m_2}\bigr)\pm \tfrac{\pi}{2}\bigr) \bmod{2\pi},& m_2\neq 0\\
	-\arccos(\mp \sign(m_1)),& m_2=0
	\end{cases};
\end{equation*}
\item two solutions if $|G(m,\eta,m_0)|<\sqrt{ m_1^2+m_2^2 }$.
\end{itemize}
\subsubsection{Spectral symmetries}
The three fundamental symmetries  impose additional constraints on the structure of the energy bands. If $U$ is invariant with respect to $T$ then 
\begin{equation*}
	\epsilon_{U,n}(k)=\epsilon_{U,n}(-k),
\end{equation*}
and the condition~\eqref{eq:Tcond} leads indeed to $F_{U,k}(\epsilon)=F_{U,-k}(\epsilon)$. If $U$ is invariant with respect to $C$ then 
\begin{equation*}
	\epsilon_{U,n}(k)=-\epsilon_{U,-n}(-k),
\end{equation*}
and the condition~\eqref{eq:Ccond} implies that  $F_{U,k}(\epsilon)=|F_{U,-k}(-\epsilon)|$. Finally, if $U$ is invariant with respect to $S$ we have that
\begin{equation*}
	\epsilon_{U,n}(k)=-\epsilon_{U,-n}(k),
\end{equation*}
and the condition~\eqref{eq:Scond} implies that  $F_{U,k}(\epsilon)=|F_{U,k}(-\epsilon)|$.

\subsection{Zak phase}\label{sec:topological}
Let us recall that the Zak phase can be defined for \emph{isolated} energy bands. In particular, if  $\epsilon_{U,n}(k)$ is a simple eigenvalue of $h_U(k)$ associated to a single isolated band, the corresponding Zak phase is defined by
 \begin{align}\label{eq:ZUn}
Z_{U,n}= \iu \int_{-\pi}^\pi \langle u_{U,n}(k,\cdot )|\partial_k u_{U,n}(k,\cdot)\rangle_{L^2((-\frac{1}{2},\frac{1}{2});\C^2)}\,\dd k
 \end{align}
 where $\partial_k=\dd/\dd{k}$ and
\begin{align}\label{eq:Zakeig}
 u_{U,n}(k,x)=\e^{-\iu k x}\Psi_{U,n}(k,x)
\end{align}
is the eigenspinor in the so-called Bloch--Floquet--Zak representation, see Appendix~\ref{App:BerryZak} for further details. Notice that since in general we cannot derive a closed expression of $\epsilon_{U,n}(k)$ as a function of $k$, $Z_{U,n}$ will be computed numerically. By discretizing the Brillouin zone, however, the integral underlying the inner product in Eq.~\eqref{eq:ZUn} can be computed analytically in terms of $\epsilon_{U,n}(k)$, see Eq.~\eqref{eq:Si} in Appendix~\ref{App:BerryZak}. Unless differently stated, the numerical value of $Z_{U,n}$ will always be given in the range $[0,2\pi)$. In the following we discuss the role of the Zak phase for the symmetry classes D, BDI and AIII.

\subsubsection{Class D}
The set $\mathcal{U}_{\text{D}}$ consists of the one-parameter family of coupling matrices $U_C(\theta)$ with $\theta\in[-\pi,\pi)\cong\mathbb{S}^1$, see Eqs.~\eqref{eq:UD}--\eqref{eq:UC}. Charge-conjugation symmetry implies that, up to a reflection in the Brillouin zone, the energy bands are symmetric with respect to $\epsilon=0$: 
\begin{equation*}
    \epsilon_{U_C(\theta),n}(k)=-\epsilon_{U_C(\theta),-n}(-k).
\end{equation*}
By inspecting the spectral function 
\begin{equation}\label{eq:FUC}
    F_{U_C(\theta),k}(\epsilon)=\sin(\theta)\sin(k)+\epsilon\sinc(q)
\end{equation}
we deduce that there is an infinite number of energy gaps. If in particular $\theta \in \{-\pi,0\}$ the coupling conditions are impermeable, and the corresponding flat bands are given by the closed expression
\begin{equation*}
	\epsilon_{U_C(-\pi),n}=\epsilon_{U_C(0),n}=\begin{cases}
        \pm\sqrt{n^2\pi^2+m^2},&n\in \Z^*\\
        0,&n=0
    \end{cases}.
\end{equation*}
If $\theta\notin \{-\pi,0\}$ the energy bands can be determined only numerically by finding the real roots of the transcendental equation $F_{U_C(\theta),k}(\epsilon)=0$. In any case from Eq.~\eqref{eq:FUC} we can still conclude that the energy gaps close only in the massless limit $m\to 0$ for $\theta=\pm\tfrac{\pi}{2}$, see Fig.~\ref{fig:UCbands}.

\begin{figure}[t]
\centering
\includegraphics{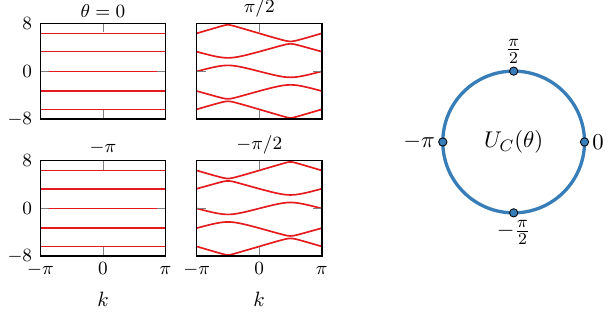}
    \caption{Energy bands close to zero for  the couplings $U_C(\theta)$ in class D, with $m=1$. The energy gaps never close, therefore the set $\mathcal{U}_{\text{D}}$ consists of a single connected region with no topological phase transitions.}
    \label{fig:UCbands}
\end{figure}

Accordingly, we expect no topological phase transition in class D, as we can connect any two elements of $\mathcal{U}_\text{D}$ by a continuous transformation, leading to a deformation of the corresponding Hamiltonians preserving charge-conjugation symmetry and, at the same time, not closing the energy gaps. 
In Fig.~\ref{fig:Zakphase1} we plot the Zak phase $Z_{U_C(\theta),n}$ for $|n|\le 2$: as it turns out, the Zak phase does not assume a quantized value as a function of $\theta$, and it therefore does not catch any topological information for this class.
 
\begin{figure}[hb]
\centering
\includegraphics{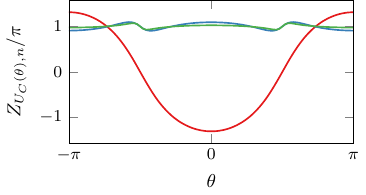}
    \caption{Zak phase $Z_{U_C(\theta),n}$ for the class D, with $n=0$ (red), $n=\pm1 $ (blue) and  $n=\pm 2 $ (green); some values of $Z_{U_C(\theta),0}$ have been shifted by $2\pi$ in order to show a continuous curve.}
    \label{fig:Zakphase1}
\end{figure}

\subsubsection{Class BDI}
The set $\mathcal{U}_{\text{BDI}}$ consists of the one-parameter family of coupling matrices  $U_{CS}(\theta)$ with $\theta\in[-\pi,\pi)\cong\mathbb{S}^1$, see Eqs.~\eqref{eq:UBDI}--\eqref{eq:UCS}. The invariance with respect to all the three fundamental symmetries implies that the energy bands are symmetric with respect to $\epsilon=0$,
\begin{equation*}
    \epsilon_{U_{CS}(\theta),n}(k)=-\epsilon_{U_{CS}(\theta),-n}(k)
\end{equation*}
and with respect to a reflection in the Brillouin zone, that is
\begin{equation*}
    \epsilon_{U_{CS}(\theta),n}(k)=\epsilon_{U_{CS}(\theta),n}(-k).
\end{equation*}
By inspecting the spectral function 
\begin{equation*}
F_{U_{CS}(\theta),k}(\epsilon)=\sin(\theta)\cos(k)+\cos(q)-m\cos(\theta)\sinc(q)
\end{equation*}
we observe that there are infinitely many energy gaps except for $\theta=\pm\tfrac{\pi}{2}$, see Fig.~\ref{fig:UCSbands}. For the latter values of $\theta$ there are only two energy bands separated by a central \emph{mass gap} $(-m,m)$. Indeed, the matrices $U_{CS}(-\tfrac{\pi}{2})=U_{\text{pp}}(0)$ and $U_{CS}(\tfrac{\pi}{2})=U_{\text{pp}}(\pi)$ correspond respectively to the anti-periodic and periodic coupling conditions introduced in Section~\ref{sec:coupcond}: in these cases the Hamiltonian is unitarily equivalent to the free Dirac operator $H$, and thus
\begin{align*}
\spec\bigl(H_{U_{CS}(\pm\frac{\pi}{2})}\bigr)=\spec(H)=(-\infty,m]\cup[m,+\infty).
\end{align*}
In particular we find the closed expressions
\begin{gather*}
    \epsilon_{U_{CS}(-\frac{\pi}{2}),\pm n}(k)=\pm\sqrt{\bigl(2\bigl\lfloor \tfrac{n}{2}\bigr\rfloor\pi-(-1)^n |k|\bigr)^2+m^2},\\
    \epsilon_{U_{CS}(\frac{\pi}{2}),\pm n}(k)=\pm\sqrt{\bigl(\bigl(\bigl\lfloor \tfrac{n-1}{2}\bigr\rfloor+1\bigr)\pi+(-1)^n |k|\bigr)^2+m^2},
\end{gather*}
where $n\in\N^\ast\coloneq\N \setminus \{0\}$ while $\lfloor x \rfloor$ is the \emph{floor} of $x$, i.e.\ the greatest integer less than or equal to $x$. 

\begin{figure}[tp]
\centering
\includegraphics{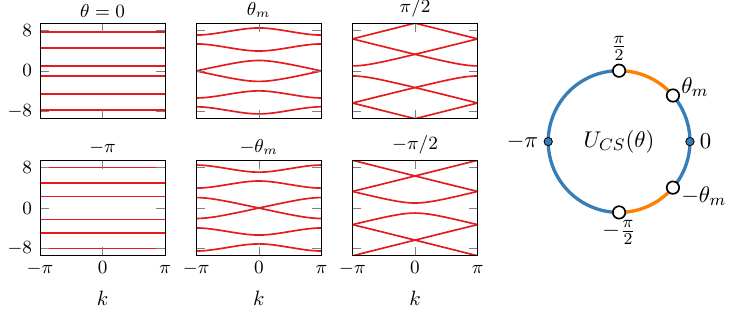}
   \caption{Energy bands close to zero for  the couplings $U_{CS}(\theta)$ in class BDI, with $m=1$. The central energy gap closes for $\theta=\pm\theta_m$, whereas all the other gaps close for $\theta=\pm\tfrac{\pi}{2}$. The set $\mathcal{U}_{\text{BDI}}$ is thus partitioned into four connected regions (represented by alternating colors in the right diagram) where no gap of the corresponding Hamiltonian closes.}
    \label{fig:UCSbands}
\end{figure}

The central energy gap, on the other hand,  can close only if there is a zero eigenvalue of $h_U(k)$. If $\theta\in\{-\pi,0\}$ the coupling conditions are impermeable, and since $F_{U_{CS}(-\pi),k}(0)=\e^{m}$ and $F_{U_{CS}(0),k}(0)=\e^{-m}$ there are no zero-energy flat bands. If instead $\theta\notin\{-\pi,0\}$, by following Section~\ref{sec:zeromodes} we compute
 \begin{align*}
  G(m,\tfrac{\pi}{2},\cos(\theta))&=\cosh(m)-\cos(\theta) \sinh(m)\\
  &=|\sin(\theta)|\cosh\bigl(\arctanh(\cos(\theta))-m\bigr),
 \end{align*}
and since $\sqrt{m_1^2+m_2^2}=|\sin(\theta)|$ we conclude that there is a zero eigenvalue only if $\theta=\pm\theta_m$, where 
\begin{align}\label{eq:thetam}
\theta_m=\arccos(\tanh(m))\in (0,\tfrac{\pi}{2}).
\end{align}
Notice that the endpoints $\theta_m=0$ and $\theta_m=\tfrac{\pi}{2}$ can only be reached in the limits $m\to+\infty$ and $m\to 0$, respectively.

By collecting the above spectral properties we identify four \emph{gapped regions} where no gap closes for the Hamiltonians with coupling in the set $\mathcal{U}_{\text{BDI}}$, depicted in the right diagram of Fig.~\ref{fig:UCSbands}. Each region corresponds to a different topological phase, and is characterized by a different value of the topological invariant: as it turns out for this symmetry class the Zak phase is indeed (numerically) constant within each gapped region, and  it is quantized in units of $\pi$:
\begin{align}\label{eq:ZUCS}
Z_{U_{CS}(\theta),n}=\begin{cases}
0, &n=\pm1,\,\theta\in(-\tfrac{\pi}{2},-\theta_m)\cup(\theta_m,\tfrac{\pi}{2})\\
\pi , &n=\pm1,\, \theta\in [-\pi,-\tfrac{\pi}{2})\cup(-\theta_m,\theta_m)\cup(\tfrac{\pi}{2},\pi)\\
\pi , &|n|>1,\,\theta\neq \pm\tfrac{\pi}{2}
\end{cases}.
\end{align}
We plot $Z_{U_{CS}(\theta),n}$ for $|n|\le 3$ in  Fig.~\ref{fig:Zakphase2}.

\begin{figure}
\centering
\includegraphics{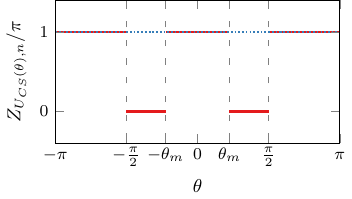}
    \caption{Zak phase $Z_{U_{CS(\theta),n}}$ for the class BDI, with $n=\pm 1$ (red) and $n=\pm2, \pm 3$ (dotted blue). }
    \label{fig:Zakphase2}
\end{figure}

\subsubsection{Class AIII}
The set $\mathcal{U}_{\text{AIII}}$ consists of the two-parameter family of coupling matrices $U_{S}(\theta,m_2)$ with $(\theta,m_2)\in[-\pi,\pi)\times [-1,1]$, see Eqs.~\eqref{eq:UAIII}--\eqref{eq:US}. Notice that  the matrices $U_S(\theta,\pm1)=\mp\sigma_y$ are independent of $\theta$, and thus the parameter space of $\mathcal{U}_{\text{AIII}}$ is in bijection with the sphere $\mathbb{S}^2$:
\begin{align*}
\Bigl\{\Bigl(\sqrt{1-m_2^2}\cos(\theta),\sqrt{1-m_2^2}\sin(\theta),m_2\Bigr)\in\R^3:(\theta,m_2)\in [-\pi,\pi)\times [-1,1]\Bigr\},
\end{align*}
see the left diagram in Fig.~\ref{fig:USparameter}. As we explained in Section~\ref{sec:symmetryclass}, the set $\mathcal{U}_{\text{AIII}}$ contains a continuous embedding of $\mathcal{U}_{\text{BDI}}$, and this deformation preserves (only) the invariance with respect to the chiral symmetry. As a consequence, the Hamiltonians in the classes BDI and AIII share some spectral properties. For $U\in\mathcal{U}_{\text{AIII}}$ the energy bands are still symmetric with respect to $\epsilon=0$, 
\begin{equation*}
    \epsilon_{U_{S}(\theta,m_2),n}(k)=-\epsilon_{U_{S}(\theta,m_2),-n}(k)\,.
\end{equation*}
Also in this case there are infinitely many energy gaps unless $\theta=\pm\tfrac{\pi}{2}$, that is when the matrices $U_{CS}(\theta, m_2)$ coincide with the pseudo-periodic conditions $U_{\text{pp}}(\alpha)$, and all the energy gaps but the central one close. Moreover from the spectral function
\begin{equation*}
\begin{aligned}
    F_{U_{S}(\theta,m_2),k}(\epsilon)&=\sqrt{1-m_2^2}\sin(\theta)\cos(k)+m_2\sin(k) \\
    &\quad +\cos(q)-m\sqrt{1-m_2^2}\cos(\theta)\sinc(q)
\end{aligned}
\end{equation*}
we deduce that there is a zero eigenvalue of $h_U(k)$ closing the central gap if 
\begin{equation*}
    \cosh(m) -\sqrt{1-m_2^2}\cos(\theta)\sinh(m)=\pm \sqrt{1-(1-m_2^2)\cos(\theta)^2}.
\end{equation*}  
The above conditions can only hold with the positive sign, and are thus equivalent to the equation
\begin{equation}\label{eq:locus}
 \sqrt{1-m_2^2}\cos(\theta)=\tanh(m),
\end{equation}  
which gives the locus where the central energy gap closes. By solving for $\theta$ we find
\begin{align*}
\tilde{\theta}_m(m_2)=\arccos\Bigl(\tfrac{\tanh(m)}{\sqrt{1-m_2^2}}\Bigr)\qquad \text{if}\qquad |m_2|\le \sqrt{1-\tanh^2(m)},
\end{align*}
where $\tilde{\theta}_m(0)=\theta_m$ coincides with the value found in Eq.~\eqref{eq:thetam} for $\mathcal{U}_{\text{BDI}}$.

\begin{figure}[tp]
\centering
\includegraphics{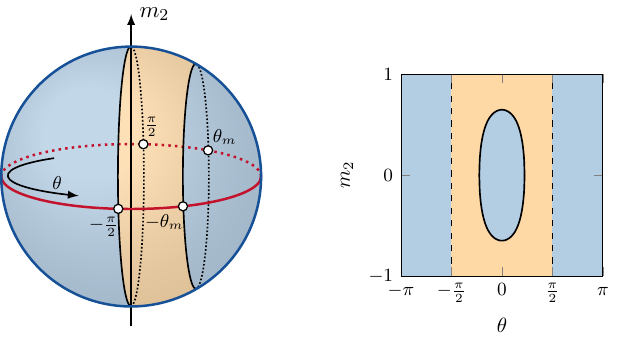}
    \caption{(Left) $\mathbb{S}^2$ parameter space of the set $\mathcal{U}_{\text{AIII}}$, where the red line at the equator $m_2=0$ represents the couplings $U_S(0,\theta)=U_{CS}(\theta)$, while the shaded regions highlight the three gapped regions~\eqref{eq:hemisph}--\eqref{eq:cap}. (Right) Phase diagram of the (quantized) Zak phase $Z_{U_{S}(\theta,m_2),\pm 1}$, with  $Z=\pi$ in blue and $Z=0$ in orange; the black solid curve and the dashed lines represent respectively the small circle  of  $\mathbb{S}^2$ where the central gap closes, see Eq.~\eqref{eq:locus}, and the great circle at $\theta=\pm\tfrac{\pi}{2}$ where all the other gaps close.}
    \label{fig:USparameter}
\end{figure}

By excluding the coupling conditions for which at least one energy gap closes, the set $\mathcal{U}_{\text{AIII}}\cong \mathbb{S}^2$ is partitioned into three gapped regions:
a hemisphere
\begin{equation}\label{eq:hemisph}
	\mathcal{U}_{\text{AIII}}^{\text{hem}} \coloneq \{U_{S}(\theta,m_2): |\theta|> \tfrac{\pi}{2}\},
\end{equation}
a spherical zone
\begin{equation}\label{eq:zone}
	\mathcal{U}_{\text{AIII}}^{\text{sz}} \coloneq \Bigl\{U_{S}(\theta,m_2) : 0<\sqrt{1-m_2^2}\cos(\theta)<\tanh(m)\Bigr\}
\end{equation}
and a spherical cap
\begin{equation}\label{eq:cap}
	\mathcal{U}_{\text{AIII}}^{\text{sc}} \coloneq \Bigl\{U_{S}(\theta,m_2) : \sqrt{1-m_2^2}\cos(\theta)>\tanh(m)\Bigr\}\,.
\end{equation} 
These regions represent three different topological phases characterized by the following values of the Zak phase:
\begin{gather}
Z_{U,n}=\begin{cases}
0, &n=\pm1\,,\; U \in \mathcal{U}_{\text{AIII}}^{\text{sz}}\,, \\
\pi, &n=\pm1\,,\; U \in \mathcal{U}_{\text{AIII}}^{\text{hem}} \cup \mathcal{U}_{\text{AIII}}^{\text{sc}}\,, \\
\pi, & |n|> 1\,,\;  U = U_{S}(\theta,m_2) \in \mathcal{U}_{\text{AIII}}\,, \; \theta\neq\pm \tfrac{\pi}{2}\,,
\end{cases}
\end{gather}
which we plot in the right panel of Fig.~\ref{fig:USparameter} for $n=\pm 1$. We remark that this result is consistent with the one already found for class BDI. Let us recall that although the  two  regions
\begin{align*}
\{U_{CS}(\theta) : \theta\in  (-\tfrac{\pi}{2},-\theta_m)\},&& 
\{U_{CS}(\theta): \theta\in  (\theta_m,\tfrac{\pi}{2})\},
\end{align*}
in the set $\mathcal{U}_{\text{BDI}}$ are characterized by equal topological indices (see Eq.~\eqref{eq:ZUCS}), the corresponding Hamiltonians cannot be connected by any continuous transformation not closing the energy gaps and preserving at the same time both charge-conjugation and chiral symmetries, and are thus associated to  two different topological phases. That notwithstanding, they can be connected by a continuous transformation not closing the gaps and preserving \emph{only} the chiral symmetry, and they are thus associated to the same topological phase within the larger class AIII.

\section{Bulk-boundary correspondence} \label{sec:BulkBoundary}
The BBC is usually expressed as the equivalence between two topological indices, characterizing respectively the infinitely extended system without boundaries and the half-infinite system obtained by truncating the first with a sharp boundary~\cite{MoSh11, ShShLu11, KaHa13, XiZhCh14, JuTa25}. For one-dimensional models, the  bulk index is typically given by the Zak phase associated to the $n$-th energy band~\cite{FuKa06, XiChNi10, KaHa13, XiZhCh14, Guo16, AtAiBa13, deJRuLe14, RhBeBa17, Res94, CaFu21}, whereas the boundary index is defined in the truncated system as difference between the number of edge states (i.e.\ normalizable eigenstates that are localized near the boundary) appearing below the $n$-th bulk energy band and the number of edge states above the same energy band~\cite{Hat93, GrLe06, TaDeVe20, GrJuTa21}.  

In continuum models, however, the truncation is not unique: both the position of the edge and the boundary conditions imposed at the edge are part of the definition of the half-infinite system. Recent studies have highlighted how the BBC is affected by the details of the truncation~\cite{MaKuIm10, DeUlMo11}, possibly leading to a \emph{violation} of the correspondence in continuum systems~\cite{TaDeVe20,GrJuTa21, RoTa24, BaCoTr24, JuTa25, GrTa25}. In this section we analyze in detail how the boundary index defined above in terms of edge states depends on two parameters characterizing all the possible truncations of the gDKP model, and how this boundary index relates to the bulk one given by the Zak phase. After defining in Section~\ref{sec:truncation} the truncated Hamiltonian and the boundary spectral function, in Section~\ref{sec:symmetriccond} we discuss the BBC for both the symmetry classes BDI and AIII but considering only \emph{symmetry-preserving} truncations. In Section~\ref{sec:stability}, focusing on the BDI class, we then discuss the stability of the correspondence with respect to arbitrary truncations.

\subsection{Truncated Hamiltonian}\label{sec:truncation}
By letting $d\in[0,1)$ denote the position of the (zero-dimensional) edge, in the following we consider the gDKP model truncated on the half-line $\R_d=(d,+\infty)$. In order for the truncated Hamiltonian to be self-adjoint, we must impose a suitable boundary condition at the edge: by recalling that the most general decoupling boundary conditions at $x=d^+$ are given by the chiral conditions introduced in Section~\ref{sec:coupcond}, namely
\begin{equation}\label{eq:alphacond}
    \cos(\tfrac{\alpha}{2})\chi(d^+)=\iu\sin(\tfrac{\alpha}{2})\phi(d^+)
\end{equation}
with $\alpha\in [-\pi,\pi)$, we define the \emph{truncated Hamiltonian} $\mathcal{H}_{U}(d,\alpha)$ as the differential operator having the same expression of $H_U$ in Eq.~\eqref{eq:HU} and domain
\begin{align*}
    \mathfrak{D}(\mathcal{H}_{U}(d,\alpha))=\{\Psi \in H^1(\R_d\setminus\N^\ast;\C^2) &: \cos(\tfrac{\alpha}{2})\chi(d^+)=\iu\sin(\tfrac{\alpha}{2})\phi(d^+),
    \nonumber\\ 
    &\quad\Psi_{-}(n)=U\Psi_{+}(n)\,\forall\, n\in\N^\ast\}.
\end{align*}
The role of the truncation parameters $d$ and $\alpha$ is shown in Fig.~\ref{fig:halfcomb}. Notice that the condition~\eqref{eq:alphacond} determines $\Psi(d^+)$ up to a factor $\mathfrak{c}\in \C$, that is
\begin{equation}\label{eq:Psichiral}
    \Psi(d^+)=\mathfrak{c} \mat{-\iu \cos(\tfrac{\alpha}{2})\\ \sin (\tfrac{\alpha}{2})}.
\end{equation}

\begin{figure}[t]
\centering
\includegraphics{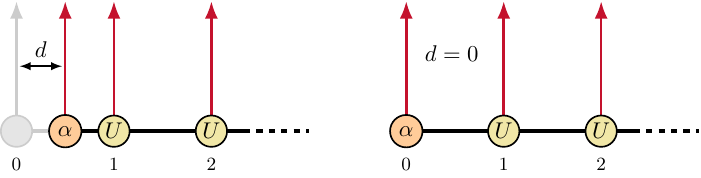}
\caption{Different truncations of the gDKP model, leading to the Hamiltonian $\mathcal{H}_{U}(d,\alpha)$ depending on the position $d$ of the edge and on the parameter $\alpha$ characterizing the boundary conditions at the edge.}
\label{fig:halfcomb}
\end{figure}

For this system, the BBC can be stated as the equivalence between $Z_{U,n}/\pi$, the normalized Zak phase associated to the $n$-th energy band of $H_U$, and the difference
\begin{align}\label{eq:NbNa}
N_{U,n}^{\text{b}}(d,\alpha)-N_{U,n}^{\text{a}}(d,\alpha)
\end{align}
between the number of edge states of $\mathcal{H}_{U}(d,\alpha)$ appearing respectively below and above the $n$-th energy band of $H_U$. 
Notice that the Zak phase actually depends on the choice of origin of the real space unit cell~\cite{AtAiBa13, deJRuLe14, RhBeBa17}.\footnote{The \emph{relative} Zak phase between two different topological phases is however independent of the unit cell convention~\cite{FuKa06, AtAiBa13, deJRuLe14}.} Although this choice is in principle arbitrary, and different conventions are connected by unitary transformations, there is usually a natural choice of the origin that is related to the presence of inversion symmetry within the unit cell. We followed this convention in Section~\ref{sec:bulk}, placing the point interaction at the center of the unit cell. For the BBC to hold, however, it is essential to connect the unit cell convention and the position of the edge appropriately~\cite{KaHa13}. In our case, this amounts to consider $d=\tfrac{1}{2}$: this case will be analyzed in Section~\ref{sec:symmetriccond}, while other values of $d$ will be discussed in Section~\ref{sec:stability}.

\subsubsection{Edge states}
The edge states can be determined by exploiting a \emph{transfer matrix} approach~\cite{Blu94, DwCh16, WiToKi25, TaDe15}.
The transfer matrix $T_U(\epsilon,d)$ associated to $H_U$ is defined by the relation
\begin{equation*}
    \Psi_\epsilon((n+d+1)^+)=T_{U}(\epsilon,d) \Psi_\epsilon((n+d)^+)
\end{equation*}
for any $n\in\Z$ and $d\in [0,1)$, where $\Psi_\epsilon(x)$ is any (generalized) eigenspinor of $H_U$ with energy $\epsilon$. For permeable conditions, the transfer matrix is given by
\begin{align}
T_{U}(\epsilon,d)=P(\epsilon,d)D_UP(\epsilon,1-d)
\end{align}
where
\begin{gather}
P(\epsilon,d)=\cos(q d)I+\iu d\sinc(q d)(\epsilon\sigma_x+\iu m\sigma_y),\\
 D_U=\frac{1}{m_1-\iu m_2}\mat{-\sin(\eta) -m_3&  \iu(\cos(\eta)+m_0) \\  \iu(\cos(\eta)-m_0) &-\sin(\eta)+m_3},
\end{gather}
see Appendix~\ref{sec:transfermatrix} for details. In this context, $\epsilon$ is in the spectrum of $H_U$ if it satisfies the well-known energy band condition $|\tr(T_U(\epsilon,d))|\le 2$, which in our case leads to the condition
\begin{equation*}
    |\cos(q)\sin(\eta)+\sinc(q)(\epsilon\cos(\eta)-mm_0)|\le \sqrt{m_1^2+m_2^2}.
\end{equation*}
Let us remark that this result is consistent with our approach based on the spectral function, as the equation $F_{U,k}(\epsilon)=0$ implies the above condition.

For what concerns the edge states of $\mathcal{H}_{U}(d,\alpha)$, let us observe that since $|\det(T_U(\epsilon,d))|=1$, $T_U(\epsilon,d)$ has two eigenvalues  $\lambda_{U,\pm}(\epsilon,d)$, reciprocal up to a phase, with $|\lambda_{U,-}(\epsilon,d)|\le 1$ and $|\lambda_{U,+}(\epsilon,d)|\ge 1$. Denoting by $v_{U,\pm}(\epsilon,d)$ the  right eigenvectors (in general $T_U(\epsilon,d)$  is not a normal matrix), we have the decomposition
\begin{equation*}
    \Psi(d^+)=\mathfrak{c}_+ v_{U,+}(\epsilon,d)+\mathfrak{c}_- v_{U,-}(\epsilon,d).
\end{equation*}
A (normalizable) edge state will thus be allowed if  $\Psi(d^+)$ belongs to the span of the \emph{decaying eigenvector}  $v_{U,-}$, or equivalently if
\begin{equation}\label{eq:decayingspan}
    w_{U,+}(\epsilon,d) \Psi(d^+)=0
\end{equation}
where $w_{U,+}(\epsilon,d)$ is the left eigenvector of $T_U(\epsilon,d)$ associated to $\lambda_{U,+}(\epsilon,d)$, namely the row vector satisfying  the relations
\begin{align*}
    w_{U,+} (\epsilon,d)T_U(\epsilon,d)=\lambda_{U,+}(\epsilon,d)  w_{U,+}(\epsilon,d),&&w_{U,+}(\epsilon,d)v_{U,-}(\epsilon,d)=0.
\end{align*}
By combining Eqs.~\eqref{eq:Psichiral} and~\eqref{eq:decayingspan} we can define the following \emph{boundary spectral function}:
\begin{equation}
    \mathcal{F}_{U}(\epsilon,d,\alpha)= w_{U,+}(\epsilon,d) \mat{-\iu \cos(\tfrac{\alpha}{2})\\ \sin (\tfrac{\alpha}{2})}.
\end{equation}
Hence, by construction, $\mathcal{H}_{U}(d,\alpha)$ has an edge state of energy $\epsilon$ if and only if $\epsilon$ is in an energy gap of $H_{U}$ and if $\mathcal{F}_{U}(\epsilon,d,\alpha)=0$.

\subsection{Symmetry-preserving truncations}\label{sec:symmetriccond}
We henceforth restrict our considerations to the coupling matrices $U_S(\theta,m_2)$ associated to the AIII class. Recall that the BDI class can be seen as a subset of the latter, since $U_{CS}(\theta)=U_{S}(\theta,0)$.  Observe that if $\Psi(x)$ satisfies the chiral boundary condition in Eq.~\eqref{eq:alphacond}, then both $C\Psi(x)=\sigma_x \overline{\Psi(x)}$ and $S\Psi(x)=-\sigma_y \Psi(x)$ satisfy the condition
\begin{equation*}
    \sin(\tfrac{\alpha}{2})\chi(d^+)=\iu\cos(\tfrac{\alpha}{2})\phi(d^+),
\end{equation*}
and this implies the following anti-commutation relations:
\begin{gather}
 S\mathcal{H}_{U_{S}(\theta,m_2)}(d,\alpha)S^{-1}=-\mathcal{H}_{U_{S}(\theta,m_2)}(d,\pi-\alpha), \\ 
 C\mathcal{H}_{U_{CS}(\theta)}(d,\alpha)C^{-1}=-\mathcal{H}_{U_{CS}(\theta)}(d,\pi-\alpha).
\end{gather}
The above relations tell us that not all the boundary conditions at the edge respect the chiral and charge-conjugation symmetries of the corresponding bulk Hamiltonian. In particular, we find that $\mathcal{H}_{U_{CS}(\theta)}(d,\alpha)$ and $\mathcal{H}_{U_{S}(\theta,m_2)}(d,\alpha)$ are respectively in the symmetry classes BDI and AIII if and only if we restrict to the values $\alpha=\pm\tfrac{\pi}{2}$.

We compute the transfer matrix $T_{U_S(\theta,m_2)}(\epsilon,d)$ and the left eigenvector $w_{U_S(\theta,m_2),+}(\epsilon,d)$ with the help of Mathematica,  obtaining a long but analytical expression for the boundary spectral function $\mathcal{F}_{U_S(\theta,m_2)}(\epsilon,d,\alpha)$. In Fig.~\ref{fig:boundary} we plot the edge spectrum of $\mathcal{H}_{U_{S}(\theta,m_2)}(d,\alpha)$ for $d=\tfrac{1}{2}$ and $\alpha=\pm\tfrac{\pi}{2}$, showing the edge states between the energy bands of $H_{U_{S}(\theta,m_2)}$. For these values of $d$ and $\alpha$ we find that the BBC holds up to a sign, that is we have that
\begin{align}\label{eq:BBC1}
\frac{1}{\pi} Z_{U_{S}(\theta,m_2),n}=s_{n,\theta,\pm \tfrac{\pi}{2}}\bigl(
     N_{U_S(\theta,m_2),n}^{\text{b}}(\tfrac{1}{2},\pm\tfrac{\pi}{2})-N_{U_S(\theta,m_2),n}^{\text{a}}(\tfrac{1}{2},\pm\tfrac{\pi}{2})\bigr)
\end{align}
for any $n\in\Z^\ast$, where $s_{n,\theta,\alpha}\in\{-1,1\}$ is given by
\begin{align}\label{eq:sign}
s_{n,\theta,\alpha}=(-1)^{n+1}\sign(n)\sign(\alpha)\sign(|\theta|-\tfrac{\pi}{2}).
\end{align}

\begin{figure}[tb]
\centering
\includegraphics{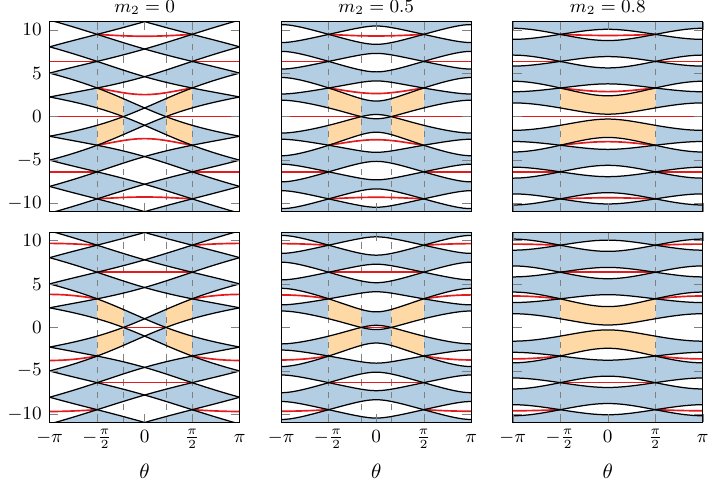}
    \caption{Edge spectrum of $\mathcal{H}_{U_{S}(\theta,m_2)}(\tfrac{1}{2},\alpha)$ for $\alpha=\tfrac{\pi}{2}$ (top row) and $\alpha=-\tfrac{\pi}{2}$ (bottom row); the edge states are shown in red, while the blue and orange regions represent respectively the bulk energy bands with $Z=\pi$ and $Z=0$.}
    \label{fig:boundary}
\end{figure}

\subsection{Non-symmetric truncations}\label{sec:stability}
We extend our analysis by discussing the stability of the BBC for non-symmetric truncations,  i.e.\ we now consider arbitrary values of both the parameters $d$ and $\alpha$. For simplicity we will focus just on the one-parameter family of coupling conditions $U_{CS}(\theta)$, that is on the BDI class. 
We start by fixing again $d=\tfrac{1}{2}$ but considering any $\alpha\in[-\pi,\pi)$. In the top row of Fig.~\ref{fig:phase1} we show the phase diagram of the number of edge states of $\mathcal{H}_{U_{CS}(\theta)}(\tfrac{1}{2},\alpha)$ in the central energy gap and in the first two positive gaps. As it turns out, for any $\alpha\notin \{-\pi,0\}$ the BBC holds again up to a sign, that is we have that
\begin{align}\label{eq:BBC2}
\frac{1}{\pi} Z_{U_{CS}(\theta),n}=s_{n,\theta,\alpha}\bigl(
     N_{U_{CS}(\theta),n}^{\text{b}}(\tfrac{1}{2},\alpha)-N_{U_{CS}(\theta ),n}^{\text{a}}(\tfrac{1}{2},\alpha)\bigr)
\end{align}
for any $n\in\Z^\ast$, where $s_{n,\theta,\alpha}$ is the same of Eq.~\eqref{eq:sign}. For $\alpha\in\{-\pi,0\}$ we find instead that the edge states coalesce with the bulk energy bands, and for $d=\tfrac{1}{2}$ we thus observe two isolated violations of the BBC. 

\begin{figure}[tb]
\centering
\includegraphics{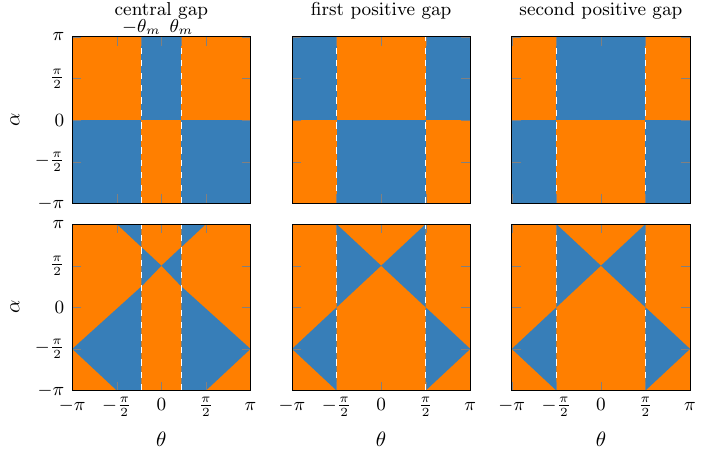}
\caption{Phase diagram of the number $N$ of edge states of $\mathcal{H}_{U_{CS}(\theta)}(d,\alpha)$ for $d=\tfrac{1}{2}$ (top row) and $d=0$ (bottom row), with $N=0$ in blue and $N=1$ in orange.}
\label{fig:phase1}
\end{figure}
 
We now consider what happens if we simultaneously change the origin of the unit cell and the edge position $d$. In order to obtain a consistent result, the edge states of $\mathcal{H}_U(d,\alpha)$ must be compared with the bulk Zak phase associated to the translated unit cell $[-1+d,d)$,  which is given (modulo $2\pi$) by
\begin{equation}
\tilde{Z}_{U,n}(d)=Z_{U,n}-\pi(1-2d)
\end{equation}
where $Z_{U,n}=\tilde{Z}_{U,n}(\tfrac{1}{2})$, see Eq.~\eqref{eq:Zd} in Appendix~\ref{sec:Zak}. Notice that even tough $\tilde{Z}_{U_{CS}(\theta),n}(d)/\pi$ happens to be an integer only for $d=\tfrac{1}{2}$ and $d=0$, the relative Zak phase between two topological phases (that is, between two values of $\theta$) is always quantized in units of $\pi$. In the bottom row of Fig.~\ref{fig:phase1} we show the same edge states phase diagram but for  $d=0$. In this case we observe that the BBC is consistently violated for any $\alpha\neq\pm\tfrac{\pi}{2}$. 

We conclude our analysis by plotting in Fig.~\ref{fig:phase2} the edge states phase diagram as a function of $d$ and fixing $\alpha=\tfrac{\pi}{2}$ (but notice that for $\alpha=-\tfrac{\pi}{2}$ we obtain analogous results). In this case we find that the edge states are insensitive to $d$ only in the central gap, where the symmetry-preserving boundary conditions with $\alpha=\pm\tfrac{\pi}{2}$ force the edge states to have zero energy.

\begin{figure}[tb]
\centering
\includegraphics{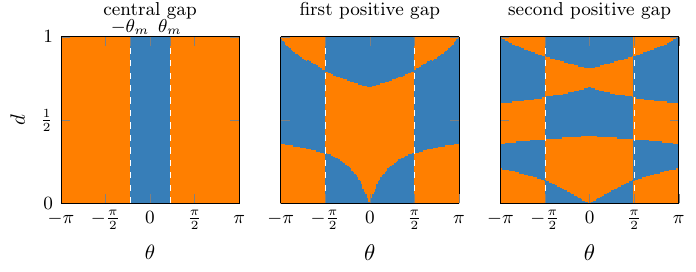}
    \caption{Phase diagram of the number $N$ of edge states of $\mathcal{H}_{U_{CS}(\theta)}(d,\tfrac{\pi}{2})$, with $N=0$ in blue and $N=1$ in orange.}
    \label{fig:phase2}
\end{figure}

\section{Discussion and outlook}
As we have shown, the gDKP Hamiltonian $H_U$ displays a rich topological structure, and the coupling matrix $U \in \UU(2)$ can be used as a parameter to explore different phases. Indeed, depending on the form of $U$, all AZC symmetry classes whose symmetries square to the identity can be exhibited as instances of this model. Through our analysis, we have thus been able to probe the topological content of the Zak phase for one-dimensional quantum systems, both in regards to bulk properties and to the BBC.

Quantization of the Zak phase is known to be protected by inversion symmetry~\cite{Zak89, Lon18}; the Zak phase can also reproduce (in an appropriate gauge) the $\Z$-valued topological index in two-band chiral systems, like the celebrated Su--Schrieffer--Heeger model~\cite{SSH79, PrSB16, MoPe23}.
In general lattice tight-binding models that are symmetric under charge-con\-jugation and/or chiral symmetry and are spectrally gapped around zero energy, a multi-band Zak phase yields a well-defined gauge and topological invariant defined as follows~\cite{MoPe23}. Assume that the model has $m$ negative and $m$ positive energy bands. In the multi-band situation, the Zak phase should be defined through \emph{quasi-Bloch functions}~\cite{Cl64}: rather than eigenfunctions of the Hamiltonian, the quasi-Bloch functions $\{\hat{u}_n(k)\}_{1 \le n \le m}$ are eigenfunctions of the spectral projection $P_-(k)$ of the Hamiltonian onto negative energy bands (which may not be isolated among themselves, but only from the positive ones). 
The charge-conjugation or chiral symmetry operator can be employed to extend this set of quasi-Bloch functions to a basis $\{\hat{u}_n(k)\}_{1 \le n \le 2m}$ of the whole fiber Hilbert space, accounting for all degrees of freedom in the unit cell. Then the quantity
\begin{equation*}
    \mathcal{I} = \iu \int_{-\pi}^{\pi} \sum_{n=1}^{2m} \langle \hat{u}_n(k,\cdot) | \partial_k \hat{u}_n(k,\cdot) \rangle \, \dd k = 2\iu \int_{-\pi}^{\pi} \sum_{n=1}^{m} \langle \hat{u}_n(k,\cdot) | \partial_k \hat{u}_n(k,\cdot) \rangle \, \dd k
\end{equation*}
is always an integer multiple of $2\pi$, whose parity is also gauge-invariant if only symmetry-preserving local gauge transformations are allowed (compare with Appendix~\ref{App:BerryZak}). Moreover, $\mathcal{I}/2\pi \bmod 2$ reproduces the parity of the $\Z$-valued index which, as predicted by Tab.~\ref{table}, can be defined for multi-band chiral chains~\cite{GrSh18}. 
The qualitative reason behind the quantization of $\mathcal{I}$ is that, if we take into account all $2m$ energy bands, then the quasi-Bloch functions $\{\hat{u}_n(k)\}_{1 \le n \le 2m}$ can be related to the quasi-Bloch functions at $k=0$ by means of a unitary matrix $U(k)$, that is
\begin{equation} \label{eqn:hatu}
\hat{u}_a(0,\cdot)=\sum_{b=1}^{2m}[U(k)]_{a,b}\hat{u}_b(k,\cdot)
\end{equation}
for any $k \in [-\pi,\pi)$. It is then possible to prove that 
\begin{equation*}
    \sum_{n=1}^{2m} \langle \hat{u}_n(k,\cdot) | \partial_k \hat{u}_n(k,\cdot) \rangle = \overline{\det(U(k))}\, \partial_k \det(U(k)),
\end{equation*}
so, using standard complex analysis, $\mathcal{I}/2\pi$ measures the integer winding number around zero of the determinant of $k \mapsto U(k)$.

The previous discussion highlights how topological content, in the form of quantization of the Zak phase, is to be expected in general tight-binding models only when considering all (gapped) spectral bands below zero as a whole, and the object that can be topologically classified is the projection $P_-(k)$~\cite{GoMoPeRo22}. 
If only single isolated bands are considered, then the Zak phase looses in general all its topological meaning, as continuous deformations can ``unwind'' this phase by exploiting the extra dimensions in the ambient space~\cite{Pe26}. 
As these extra dimensions are in principle always accessible in continuum models, where the fiber Hilbert space in the Bloch--Floquet(--Zak) representation is infinite-dimensional, the role of the Zak phase as a topological marker is challenged~\cite{Kui65}: considering single bands means taking matrices $U(k)$ as in Eq.~\eqref{eqn:hatu} that are not unitary anymore, so that the winding number of their determinant may be undefined.

In the gDKP model analyzed in this paper, we find indeed non-quantized values of the Zak phase in class D (see Fig.~\ref{fig:Zakphase1}): hence the Zak phase is \emph{not} a topological marker for the predicted $\Z_2$-valued index from Tab.~\ref{table}. Surprisingly, the model exhibits \emph{quantized} Zak phases in the chiral classes BDI and AIII (see Figs.~\ref{fig:Zakphase2} and~\ref{fig:USparameter}), signaling that in the presence of chiral symmetry the Zak phase should be related to a more robust (relative) topological invariant for (differences of) phases. Since these findings are seemingly in contrast with the general theoretical scheme sketched above, we plan to devote further investigation to this point in future work. In particular, it is not yet clear how to devise other markers to capture the topological properties of this specific model.

For what concerns the BBC, we found that the Zak phase still detects the presence of edge states in the form of a relative boundary topological index associated to spectral gaps. However, our results show that in continuum models the presence of localized edge states depends not only on bulk topological data, but also on how the system is truncated.  
This is an additional confirmation of the fact that these edge topological indices depend on the choice of the periodicity cell, as stated in Appendix~\ref{sec:Zak}; on top of that also the boundary condition of the truncated model can influence its value.  
Specifically, the truncation of the one-dimensional system studied in this paper depends on two parameters $(d,\alpha)\in [0,1)\times [-\pi,\pi)$ characterizing respectively the position of the cut and the additional boundary conditions at the edge. For the symmetry class BDI  we established that the BBC holds as the identity
\begin{equation}
    \frac{1}{\pi}\tilde{Z}_{U,n}(d)=|N_{U,n}^{\text{b}}(d,\alpha)-N_{U,n}^{\text{a}}(d,\alpha)|,
\end{equation}
for any $n\in\Z^\ast$, when:
\begin{itemize}
    \item $d=\tfrac{1}{2}$ and $\alpha\in(-\pi,\pi)\setminus\{0\}$, see Eq.~\eqref{eq:BBC1} and Fig.~\ref{fig:phase1};
    \item $d=0$ and $\alpha=\pm\tfrac{\pi}{2}$, see Eq.~\eqref{eq:BBC1} and  Fig~\ref{fig:phase2}.
\end{itemize}
For all the other values of $d$ and $\alpha$ we observe instead a systematic violation of the correspondence. We expect analogous results to hold also for the symmetry class AIII, although for the latter we considered explicitly only the parameters $d=\tfrac{1}{2}$ and $\alpha=\pm\tfrac{\pi}{2}$ (see Fig.~\ref{fig:boundary}). Notice that (isolated) violations of the correspondence have been recently observed also in related continuum models~\cite{TaDeVe20, GrJuTa21, RoTa24, BaCoTr24, JuTa25, GrTa25}. 
It thus turns out that the BBC is very sensitive to the position $d$ of the cut. Despite this, some ``stability'' can be gained when the truncated system belongs to the same symmetry class of the corresponding bulk Hamiltonian: when $\alpha=\pm\tfrac{\pi}{2}$ also the boundary conditions at the edge are symmetric with respect to $S$ and $C$, and in this case we find that the number of (zero energy) edge states in the central gap is completely unaffected by the value of $d$ (see the left panel of Fig.~\ref{fig:phase2}). We think that this positive result can be of practical use also for more realistic models discussing the role of Majorana zero modes in the context of topological quantum computation~\cite{Ki01, SaFrNa15, SaLaAg15}.

The gDKP model provides a versatile continuum framework to explore one-dimensional topological phases beyond tight-binding descriptions. Natural extensions of this work include the incorporation of disorder or quasi-periodic point interactions, where transfer-matrix methods remain applicable and may shed light on the stability of edge states in non-periodic settings. It would also be interesting to generalize our analysis to Dirac operators acting on four-component spinors, enabling access to additional symmetry classes and allowing closer contact with experimentally relevant systems such as graphene-based structures. More generally, our results highlight the importance of boundary conditions in continuum models for topological matter and suggest that a careful treatment of self-adjoint extensions is essential for a consistent formulation of the BBC in continuum quantum systems.

\appendix

\section{Singular Dirac operator}\label{app:pointint}
In this Appendix we discuss the connection between the gDKP model $H_U$ in Eq.~\eqref{eq:HU} and the singular Dirac operator $\tilde{H}_{\vb{g}}$ with a periodic array of Dirac $\ddelta$-potentials  introduced in Eq.~\eqref{eq:Hg}. After clarifying this connection in Section~\ref{sec:relation}, we take advantage of it in Section~\ref{sec:transfermatrix} to compute the transfer matrix associated to $H_U$.

\subsection{Point interactions and coupling conditions}\label{sec:relation}
The most general singular perturbation of the free Dirac operator $H$ consisting of a single point interaction at $x=0$ is given by 
\begin{align*}
    \tilde{H}_{0,\vb{g}}=-\iu\sigma_x\frac{\dd}{\dd x}+m\sigma_z +V_{\vb{g}}\ddelta(x),&& V_{\vb{g}}=\mat{g_0+g_3 & g_1-\iu g_2 \\  g_1+\iu g_2 &  g_0- g_3},
\end{align*}
where $V_{\vb{g}}$ is the Hermitian matrix containing the couplings 
\begin{align*}
\vb{g}=(g_0,g_1,g_2,g_3)\in\R^4
\end{align*}
and $\ddelta(x)$ is the Dirac delta distribution~\cite{AlbKur97, HeTu22, CLMT23}.\footnote{With respect to the notations of~\cite{CLMT23} we have $g_0=\eta$, $g_1=\omega$, $g_2=\lambda$ and $g_3=\tau$.} On a suitable spinor $\Psi(x)$ the point interaction enforces the coupling conditions 
\begin{align}\label{eq:VBC1}
   (2\iu\sigma_x - V_{\vb{g}})\Psi(0^+)=(2\iu\sigma_x + V_{\vb{g}})\Psi(0^-), 
\end{align}
which are equivalent to the following \emph{jump-average} conditions:
\begin{equation}\label{eq:VBC2}
    V_{\vb{g}}(\Psi(0^-)+\Psi(0^+))=-2\iu\sigma_x(\Psi(0^-)-\Psi(0^+)).
\end{equation}
Since
\begin{equation*}
    \det(2\iu\sigma_x\pm V_{\vb{g}})=4+g_0^2-g_1^2-g_2^2-g_3^2 \mp 4\iu g_1
\end{equation*}
the matrices $2\iu\sigma_x\pm V_{\vb{g}}$ are not invertible when
\begin{align*}
g_1=0\qquad \text{and}\qquad g_0^2-g_2^2-g_3^2=-4,
\end{align*}
leading to the decoupling of the conditions
\begin{align*}
(2\iu\sigma_x - V_{\vb{g}})\Psi(0^+)=0,&&
(2\iu\sigma_x + V_{\vb{g}})\Psi(0^-)=0,
\end{align*}
which means that the point interaction is impermeable. Notice that the parameter $g_1$ can always be set to zero by means of a singular gauge transformation~\cite{CLMT23}.

On the other hand, the most general coupling conditions leading to a self-adjoint extension $H_{0,U}$ of $H$, when initially defined in $C_0^\infty(\R\setminus\{0\};\C^2)$, are given by
\begin{align*}
    \Psi_-(0)=U\Psi_+(0), &&\Psi_\pm(0)= \frac{1}{\sqrt{2}}\mat{ \phi(0^-) \pm \chi(0^-) \\[2pt] \phi(0^+) \mp \chi(0^+) }
\end{align*}
for $U\in\UU(2)$. In order to understand the relation between the above coupling conditions and those in Eqs.~\eqref{eq:VBC1}--\eqref{eq:VBC2}, let us notice that
\begin{gather*}
    \Psi(0^-)+\Psi(0^+)=\Lambda(\Psi_+(0)+\sigma_x\Psi_-(0)), \\
    \Psi(0^-)-\Psi(0^+)=\sigma_x\Lambda(\Psi_+(0)-\sigma_x\Psi_-(0)), 
\end{gather*}
where
\begin{equation*}
    \Lambda=\frac{1}{\sqrt{2}}\mat{1 & 1 \\ 1 & -1}=\Lambda^\dagger
\end{equation*}
is a unitary matrix. Then, by substituting $\Psi_-(0)=U\Psi_+(0)$ in Eq.~\eqref{eq:VBC2} and setting $\tilde{V}_{\vb{g}}=\Lambda V_{\vb{g}}\Lambda$  and $\tilde{U}=-\sigma_x U$ we find that 
\begin{align*}
\tilde{V}_{\vb{g}}(I-\tilde{U})=-2\iu (I+\tilde{U}),
\end{align*}
i.e.\ that $-\tfrac{1}{2}\tilde{V}_{\vb{g}}$ and $\tilde{U}$ are related by an inverse Cayley transform:
\begin{equation}\label{eq:VU}
    -\frac{1}{2} \tilde{V}_{\vb{g}}=\iu (I+\tilde{U})(I-\tilde{U})^{-1}.
\end{equation}
We recall that to any Hermitian matrix $V$ there corresponds a unique unitary matrix given by the Cayley transform $\mathcal{C}(V)=(V-\iu I)(V+\iu I)^{-1}$, with $1\notin\spec(\mathcal{C}(V))$, whereas the inverse Cayley transform of a unitary matrix $U$ can be considered only if $1\notin \spec(U)$. In our case, by using the $\UU(2)$ parametrization introduced in Eq.~\eqref{eq:Uparam}, we have that $1\in\spec(\tilde{U})$ if and only if
\begin{equation*}
    m_1=\sin(\eta).
\end{equation*}
If $m_1\neq \sin(\eta)$ we derive the one-to-one correspondence 
\begin{equation*}
    H_{0,U}=\tilde{H}_{0,\vb{g}(U)}
\end{equation*}
where the relation $\vb{g}=\vb{g}(U)$ is obtained from Eq.~\eqref{eq:VU}, and is given by
\begin{equation}\label{eq:gU}
    \mat{g_0 \\ g_1  \\ g_2  \\ g_3 }=\frac{2}{\sin(\eta)-m_1}\mat{\cos(\eta)\\ m_2\\ m_3\\ -m_0}.
\end{equation}
If $m_1=\sin(\eta)$ then different operators $H_{0,U}$ can formally correspond, in general, to a certain $\tilde{H}_{0,\vb{g}(U)}$ having infinite values of the coupling, that is with $\vb{g}(U)\in(\R\cup\{\infty\})^4$.

In order to invert Eq.~\eqref{eq:gU} it is convenient to introduce the quantity
\begin{equation*}
    \Delta=4-g_0^2+g_1^2+g_2^2+g_3^2.
\end{equation*}
We find the following relations:
\begin{itemize}
\item if $g_0\neq 0$ and $\Delta\neq0$ then
\begin{gather*}
\eta=\arctan\biggl(\frac{\Delta}{4g_0}\biggr)\bmod \pi, \\
    \mat{m_0 \\ m_1 \\ m_2 \\ m_3}=\frac{\sign(\Delta)}{\sqrt{16g_0^2+\Delta^2}}\mat{-4g_3\\ \Delta-8\\ g_1\\ g_2};
\end{gather*}
\item if $g_0=0$  and $\Delta\ge 4 $ then
\begin{gather*}
\eta=\frac{\pi}{2}, \\
    \mat{m_0 \\ m_1 \\ m_2 \\ m_3}=\frac{1}{4+g_1^2+g_2^2+g_3^2}\mat{-4g_3\\ g_1^2+g_2^2+g_3^2-4\\ g_1\\ g_2};
\end{gather*}
\item if $\Delta=0$ and $g_0^2\ge 4$ then
\begin{gather*}
\eta=0, \\
    \mat{m_0 \\ m_1 \\ m_2 \\ m_3}=\frac{1}{g_0}\mat{-g_3\\ -2\\ g_1\\ g_2}.
\end{gather*}
\end{itemize}
Observe that the quantities $16g_0^2+\Delta^2$ and $4+g_1^2+g_2^2+g_3^2$ never vanish for $\vb{g}\in \R^4$, therefore we exhausted all the possibilities. Moreover, let us notice that $\vb{g}=(0,g_1,0,0)$ corresponds to
 \begin{align*}
\eta=\frac{\pi}{2}, && m_0=m_3=0,&&  m_1=\frac{g_1^2+4}{g_1^2-4},&&m_2=\frac{g_1}{g_1^2-4},
\end{align*}
whereas the decoupling conditions $g_1=0$ and $g_0^2-g_2^2-g_3^2=-4$ lead to
\begin{align*}
m_1=m_2=0,
\end{align*}
consistently with the discussion in Section~\ref{sec:coupcond}. 
\subsection{Transfer matrix}\label{sec:transfermatrix} 
In this section we derive the expression of the transfer matrix  associated to the gDKP model $H_U$. The transfer matrix $T_U(\epsilon,d)$ is defined by the relation
 \begin{equation*}
    \Psi_\epsilon((n+d+1)^+)=T_{U}(\epsilon,d) \Psi_\epsilon((n+d)^+)
\end{equation*}
for any $n\in\Z$ and $d\in [0,1)$, where $\Psi_\epsilon(x)$ is any (generalized) eigenspinor of $H_U$ with energy $\epsilon$. Let us start by observing that if $g_1\neq 0$ and $g_0^2-g_2^2-g_3^2\neq -4$, the coupling conditions in Eq.~\eqref{eq:VBC1} can be put in the form
\begin{equation*}
    \Psi(0^+)=\tilde{D}_{\vb{g}}\Psi(0^-)
\end{equation*}
where
\begin{align*}
    \tilde{D}_{\vb{g}}&=(2\iu\sigma_x -V_{\vb{g}})^{-1}(2\iu\sigma_x+V_{\vb{g}})\\
    &=-(-\tfrac{1}{2}\sigma_x V_{\vb{g}}+\iu I)^{-1}(-\tfrac{1}{2}\sigma_x V_{\vb{g}}-\iu I)\\
    &=-\mathcal{C}(-\tfrac{1}{2}\sigma_x V_{\vb{g}})\\
   &= \mat{\frac{4-g_0^2+g_1^2+g_2^2+g_3^2-4g_2}{4+g_0^2-g_1^2-g_2^2-g_3^2 + 4\iu g_1} & \frac{-4\iu(g_0+g_3)}{4+g_0^2-g_1^2-g_2^2-g_3^2 + 4\iu g_1} \\[8pt] \frac{-4\iu(g_0-g_3)}{4+g_0^2-g_1^2-g_2^2-g_3^2 + 4\iu g_1} &\frac{4-g_0^2+g_1^2+g_2^2+g_3^2+4g_2}{4+g_0^2-g_1^2-g_2^2-g_3^2 + 4\iu g_1} }.
\end{align*}
 Since
\begin{equation*}
    \det(\tilde{D}_{\vb{g}})=\frac{\det(2\iu\sigma_x+ V_{\vb{g}})}{\det(2\iu\sigma_x- V_{\vb{g}})}=\frac{4+g_0^2-g_1^2-g_2^2-g_3^2 - 4\iu g_1}{4+g_0^2-g_1^2-g_2^2-g_3^2 + 4\iu g_1}
\end{equation*}
we have that $|\det(\tilde{D}_{\vb{g}})|=1$. Notice that in general $\sigma_x V_{\vb{g}}$ is not a Hermitian matrix, hence $\tilde{D}_{\vb{g}}$ is not a unitary matrix. By recalling the relation~\eqref{eq:VU}, we can compute the interaction part of the transfer matrix associated to $H_{0,U}$, namely
\begin{align*}
    D_U&=\tilde{D}_{\vb{g}(U)}\\
    &=-\mathcal{C}(\sigma_x \Lambda \mathcal{C}^{-1}(-\sigma_x U)\Lambda)\\
    &=\frac{1}{m_1-\iu m_2}\mat{-\sin(\eta) -m_3&  \iu(\cos(\eta)+m_0) \\  \iu(\cos(\eta)-m_0) &-\sin(\eta)+m_3}
\end{align*}
where  $\mathcal{C}^{-1}(U)=\iu(I-U)^{-1}(I+U)$ is the inverse Cayley transform of $U$. 
Now let us consider an eigenspinor $\Psi_\epsilon(x)$ of $H_{0,U}$ with eigenvalue $\epsilon\in\R$. For any $x\neq 0$, $\Psi_\epsilon(x)$ satisfies the equation 
 \begin{equation*}
     -\iu\sigma_x \Psi_\epsilon'(x)+m\sigma_z\Psi_\epsilon(x)=\epsilon\Psi_\epsilon(x)
 \end{equation*}
 which we can put in the form $\Psi_\epsilon'(x)=\iu Q(\epsilon)\Psi_\epsilon(x)$ where
 \begin{align*}
     Q(\epsilon)=\epsilon\sigma_x+\iu m\sigma_y=\mat{0 & \epsilon+m\\  \epsilon-m &0}.
 \end{align*}
 Since $Q(\epsilon)^2=q^2 I$, where $q=\e^{\iu \arg(\epsilon^2-m^2)/2}\sqrt{|\epsilon^2-m^2|}$ has been introduced in Eq.~\eqref{eq:wavenumber},
 we obtain the following expression of the free propagator associated to a path of length $0\le d\le 1$:
 \begin{equation*}
     P(\epsilon,d)=\e^{\iu d Q(\epsilon)}=\cos(q d)I+\iu d\sinc(q d)Q(\epsilon).
 \end{equation*}          
 
By combining the above results, the full transfer matrix associated to $H_U$ is given by the expression
\begin{align*}
     T_{U}(\epsilon,d)&=P(\epsilon,d)D_UP(\epsilon,1-d).
 \end{align*}
Observe that since
 \begin{align*}
     \det(D_U)=\frac{m_1+\iu m_2}{m_1-\iu m_2},&& \det(P(\epsilon,d))=\e^{\iu d \tr(Q(\epsilon))}=1,
 \end{align*}
the transfer matrix is unimodular, that is $|\det(T_{U}(\epsilon,d))|=1$.
     
\section{Eigenvalue problem of the fiber Hamiltonian}\label{app:spectral}
In this Appendix we discuss the eigenvalue equation 
\begin{align}
(h_U(k)-\epsilon_U(k))\Psi_{\epsilon}(k,x)=0
\end{align}
associated to the fiber Hamiltonian $h_U(k)$ introduced in Eq.~\eqref{eq:hU}.  The above equation  consists of two coupled differential equations, given by
\begin{align}\label{eq:eigenDirac}
	-\iu\phi_{\epsilon}'(k,x)=(\epsilon+m) \chi_{\epsilon}(k,x),&&
 	-\iu\chi_{\epsilon}'(k,x)=(\epsilon-m) \phi_{\epsilon}(k,x),
\end{align}
For a given $\epsilon\in\mathbb{R}$ the space of \emph{pseudo-periodic} solutions (that is, satisfying the relation $\Psi_{\epsilon}(k,\tfrac{1}{2})=\e^{\iu k}\Psi_{\epsilon}(k,-\tfrac{1}{2})$) in $H^1((-\tfrac{1}{2},\tfrac{1}{2})\setminus\{0\})$  is two-dimensional, and the general solution  can be written as
\begin{align*}
\Psi_{\epsilon}(k,x)=c^+\mat{\phi_{\epsilon}^{+}(k,x)\\[2pt] \chi_{\epsilon}^{+}(k,x )}+c^-\mat{\phi_{\epsilon}^{-}(k,x)\\[2pt] \chi_{\epsilon}^{-}(k,x)},
\end{align*}
where $c^\pm\in\mathbb{C}$. After inserting the above solution in the boundary values $\Psi_{\pm}(0)$ and setting
\begin{equation*}
\Psi_{\pm}(0)=A_{k,\pm}(\epsilon)\mat{
c^+ \\ c^-
}
\end{equation*}
where
\begin{equation}\label{eq:ApmD}
A_{k,\pm}(\epsilon)= \mat{ 
\phi_{\epsilon}^{+}(k,0^- ) \pm  \chi_{\epsilon}^{+}(k,0^- ) 
 & \phi_{\epsilon}^{-}(k,0^- )  \pm  \chi_{\epsilon}^{-}(k,0^- )  \\[4pt]
\phi_{\epsilon}^{+}(k,0^+ )  \mp  \chi_{\epsilon}^{+}(k,0^+ )
 &  \phi_{\epsilon}^{-}(k,0^+ ) \mp  \chi_{\epsilon}^{-}(k,0^+ )},
\end{equation}
imposing the coupling condition $\Psi_{-}=U\Psi_{+}$ is equivalent to requiring the  vanishing of the following spectral function:
\begin{align} 
F_{U,k}(\epsilon)= \det\bigl(B_k(\epsilon)-U\bigr)=0, &&
B_k(\epsilon)= A_{k,-}^{}(\epsilon)A_{k,+}^{-1}(\epsilon).
\end{align}

After determining the explicit expression of the eigenspinors in Section~\ref{eq:eigspin}, we derive the spectral function in Section~\ref{eq:spectralfun}.
\subsection{Eigenspinors}\label{eq:eigspin}
If $\epsilon\neq \pm m$ we can rearrange Eq.~\eqref{eq:eigenDirac} as
\begin{align*}
    \phi_{\epsilon}''(k,x)=-q^2 \phi_{\epsilon}(k,x),&& \chi_{\epsilon}(k,x)=-\frac{\iu}{\epsilon+m} \phi_{\epsilon}'(k,x)
\end{align*}
where $q=q(\epsilon)=\e^{\iu \arg(\epsilon^2-m^2)/2}\sqrt{|\epsilon^2-m^2|}$ has been introduced in Eq.~\eqref{eq:wavenumber}. The above equations can be easily integrated, giving the pseudo-periodic solution presented in Section~\ref{sec:Eigenspinors}:
\begin{equation*}\label{eq:ppsol}
    \Psi_{\epsilon}(k,x)=\begin{cases}
    \xi^+
        \e^{\iu q x}
        +\xi^- \e^{-\iu q x},& x<0 \\[2pt]
        \xi^+ \e^{\iu [q(x-1)+k]}
        +\xi^- \e^{-\iu [q(x-1)-k]},& x>0
    \end{cases}
\end{equation*}
where 
\begin{align*}
    \xi^\pm=c^\pm  \mat{1 \\ \pm \frac{q}{\epsilon+m}}.
\end{align*}
The coupling condition equation $\Psi_-(0)=U\Psi(0)$, that is equivalent to
\begin{align}
    M_{U,k}(\epsilon)\mat{c^-\\ c^+}=0
\end{align}
where
\begin{align}
M_{U,k}(\epsilon)=A_{-,k}(\epsilon)-UA_{+,k}(\epsilon)=\mat{m_{11} & m_{12} \\ m_{21} & m_{22}},
\end{align}
can be used to determine the coefficients $c^\pm$. By setting
\begin{align*}
    c^{+}=m_{12}, &&c^{-}=-m_{11},
\end{align*}
we find (up to a prefactor) the explicit expression 
\begin{align*}
    c^{\pm}=\pm \bigl(1\pm \tfrac{q}{\epsilon+m}\bigr)(1-\e^{\iu (k+\eta\pm q)} (\iu m_1+m_2))\mp \bigl(1\mp\tfrac{q}{\epsilon+m}\bigr)\e^{\iu \eta}(m_0+\iu m_3).
\end{align*}

\subsection{Spectral function}\label{eq:spectralfun}
In order to determine the explicit solution of the spectral function, let us first notice that by making use of the relation 
\begin{equation*}
\det(M-N)=\det(M)+\det(N) + \tr(MN) - \tr(M)\tr(N),
\end{equation*}
that holds for any pair of $2\times 2$ matrices, we obtain that
\begin{equation}\label{eq:FU}
F_{U,k}(\epsilon)=\det(B_k(\epsilon))+\det(U)+\tr(B_k(\epsilon)U)-\tr(B_k(\epsilon))\tr(U).
\end{equation}
Then by using the solution~\eqref{eq:ppsol} we find
\begin{align*}
    A_{k,\pm}(\epsilon)=\mat{ 1\pm \frac{q}{\epsilon+m} & 1\mp \frac{q}{\epsilon+m}  \\[4pt]
    \Bigl(1\mp \frac{q}{\epsilon+m}\Bigr)\e^{\iu(k-q)} & \Bigl(1\pm \frac{q}{\epsilon+m} \Bigr)\e^{\iu(k+q)} }
\end{align*}
and
\begin{equation*}
    B_k(\epsilon)=a(\epsilon)I+b(\epsilon) \sigma_k
\end{equation*}
where
\begin{gather*}
    \sigma_k=\cos(k)\sigma_x+\sin(k)\sigma_y=\mat{0 & \e^{-\iu k} \\  \e^{\iu k}  & 0}
\end{gather*}
and
\begin{align*}
    a(\epsilon)=\frac{ m \sin(q)}{\epsilon \sin(q) -\iu q \cos(q)},&&
    b(\epsilon)=\frac{ -\iu q}{\epsilon \sin(q) -\iu q \cos(q)}.
\end{align*}
By using the expression in Eq.~\eqref{eq:FU} we get
\begin{equation}\label{eq:F}
    F_{U,k}(\epsilon)=a(\epsilon)^2-b(\epsilon)^2+\det(U)-a(\epsilon)\tr(U)+b(\epsilon)\tr(U\sigma_k)
\end{equation}
and by dropping a non-vanishing multiplicative factor and using the $\UU(2)$ parametrization in Eq.~\eqref{eq:Uparam} we further arrive at the equivalent expression
\begin{equation}\label{eq:tildeF}
F_{U,k}(\epsilon)=m_1 \cos(k)+m_2\sin(k)+\cos(q)\sin(\eta)+\sinc(q)(\epsilon\cos(\eta)-mm_0).
\end{equation}
The spectral function in Eq.~\eqref{eq:tildeF} is always a real function of $\epsilon$, contrarily to the one in Eq.~\eqref{eq:F}. Let us stress that although the above spectral function has been derived by assuming that $\epsilon\neq \pm m$, by proceeding on the line of~\cite{isoboundary,isodirac} one can prove that it can be used consistently also when $\epsilon=\pm m$.

\section{Berry and Zak phases}\label{App:BerryZak}
In this Appendix, after recalling in Section~\ref{sec:Berry} the definition of the Berry phase associated to a generic parameter space, in Section~\ref{sec:Zak} we discuss how it can be adapted when the parameter space is the (one-dimensional) Brillouin zone, leading to the Zak phase.
\subsection{Berry phase}\label{sec:Berry}
Let $H(\vb{\xi})$ denote a family of Hamiltonians depending on a (vector) parameter $\vb{\xi}$. We consider a cyclic evolution across a \emph{closed} path $C$ in the parameter space, assuming that the Hamiltonians $H(\vb{\xi})$ share a common domain and admit an eigenstate $\phi(\vb{\xi})$ associated to a (simple) eigenvalue $\epsilon(\xi)$ that does not cross with other eigenvalues for all $\vb{\xi}\in C$. By denoting the endpoints of the path $C$ by $\vb{\xi}_\text{i}$ and $\vb{\xi}_\text{f}$, with $\vb{\xi}_\text{i}=\vb{\xi}_\text{f}$, we clearly have that
\begin{align}\label{eq:xi_if}
H(\vb{\xi}_\text{i})=H(\vb{\xi}_\text{f}),&&
\phi(\vb{\xi}_\text{i})=\phi(\vb{\xi}_\text{f}).
\end{align}
The (Pancharatnam--)Berry phase associated to $\phi(\vb{\xi})$ and  $C$ is defined by the loop integral of the Berry connection~\cite{Pan56, Be84,Sim83, Res94, CaFu21}:
\begin{align*}
\gamma_C= \oint_C \mathcal{A}(\vb{\xi}),&&\mathcal{A}(\vb{\xi})=\iu\langle \phi(\vb{\xi})|\vb{\nabla}_{\vb{\xi}} \phi(\vb{\xi})\rangle\cdot\dd\vb{\xi}.
\end{align*} 
Local gauge transformations in the space of parameters,
\begin{align*}
\phi(\vb{\xi})\mapsto\e^{\iu\theta(\vb{\xi})}\phi(\vb{\xi}),
\end{align*}  
are \emph{allowed} if they respect the condition in Eq.~\eqref{eq:xi_if}, which implies that $\theta(\vb{\xi})$ must satisfy
\begin{align}\label{eq:thetaif}
\theta(\vb{\xi}_\text{f})-\theta(\vb{\xi}_\text{i})\in 2\pi\Z ,
\end{align}
but can possibly be a multivalued function. Notice that while the Berry connection is gauge-dependent,
\begin{align*}
\mathcal{A}(\vb{\xi})\mapsto \mathcal{A}(\vb{\xi})+\vb{\nabla}_{\vb{\xi}} \theta(\vb{\xi})\cdot\dd\vb{\xi},
\end{align*}
the Berry phase is gauge-invariant modulo $2\pi$, 
\begin{align*}
\gamma_C\mapsto \oint_C \mathcal{A}(\vb{\xi})+\theta(\vb{\xi}_\text{f})-\theta(\vb{\xi}_\text{i}),
\end{align*}
whereas the so-called Abelian  Wilson loop $W_C =\e^{\iu\gamma_C}$ is always gauge-invariant.

In some cases, we can obtain a measurable (that is, gauge-invariant) quantity also for an \emph{open} path from $\vb{\xi}_\text{i}$ to  $\vb{\xi}_\text{f}$. Let us suppose that $H(\vb{\xi}_\text{i})$ and $H(\vb{\xi}_\text{f})$ are related by a unitary operator $\tau$:
\begin{align*}
H(\vb{\xi}_\text{i})=\tau H(\vb{\xi}_\text{f}) \tau^\dagger ,&& \phi(\vb{\xi}_\text{i})=\tau \phi(\vb{\xi}_\text{f}),
\end{align*}
Then we can still obtain a gauge-independent Berry phase (modulo $2\pi$) if the unitary operator commutes with local gauge transformations,
\begin{align*}
\e^{\iu\theta(\vb{\xi_\text{i}})}\phi(\vb{\xi}_\text{i})=
\tau (\e^{\iu\theta(\vb{\xi_\text{f}})}\phi(\vb{\xi}_\text{f}))=
\e^{\iu\theta(\vb{\xi_\text{f}})}\tau \phi(\vb{\xi}_\text{f})=
\e^{\iu\theta(\vb{\xi_\text{f}})} \phi(\vb{\xi}_\text{i}),
\end{align*}
so that Eq.~\eqref{eq:thetaif} still holds.
\subsection{Zak phase}\label{sec:Zak}
In a periodic system we can consider the Brillouin zone $[-\pi,\pi)$ (supposing  for simplicity a unit lattice spacing) as the parameter space to compute the Berry phase. In this case, indeed,  one can e.g.\ apply an electric field to cause a linear variation in $k$ across the entire Brillouin zone~\cite{Zak89, Res94, Res00, XiChNi10, CaFu21}. The initial and final parameters $k_\text{i}=-\pi$ and $k_\text{f}=\pi$ are thus connected by the reciprocal lattice vector $2\pi$, and if the system is described by the Hamiltonian $H(k)$ in the Bloch--Floquet--Zak representation (see e.g.~\cite{Res94,XiChNi10}), we have that
\begin{align*}
H(-\pi)=\tau_{2\pi} H(\pi)\tau_{2\pi}^\dagger,
\end{align*}
where the unitary operator
\begin{align*}
\tau_{k}\colon u(x)\mapsto \e^{\iu k x} u(x)
\end{align*}
commutes with the local gauge transformations $u(x)\mapsto \e^{\iu \theta(k)}u(x)$. In the Bloch--Floquet--Zak representation the eigenfunctions $u_{n}(k,x)$ of $H(k)$ satisfy the \emph{periodic gauge}
\begin{align*}
u_{n}(k+k',x)=\e^{-\iu k'x}u_{n}(k,x)
\end{align*}
for each $k'\in2\pi\Z$. We can thus define an open-path Berry phase across the Brillouin zone, which in this context is known as the Zak phase
\begin{align*}
Z_n= \iu \int_{-\pi}^\pi \langle u_{n}(k,\cdot)|\partial_k u_{n}(k,\cdot)\rangle \,\dd k,
\end{align*}
where the inner product is the one associated to the Hilbert space of the (Wigner-Seitz) unit cell. 

As is well-known, the Zak phase is not invariant under unitary translations of the unit cell. Indeed, under the translation
\begin{align}
[-\tfrac{1}{2},\tfrac{1}{2})\mapsto [-1+d,d)
\end{align}
by the quantity $\tfrac{1}{2}-d$, where $d\in [0,1)$, the eigenfunctions in the Bloch--Floquet--Zak representation change as
\begin{align*}
u_n(k,x)\mapsto \tilde{u}_{n}(k,x-\tfrac{1}{2}+d)=\e^{\iu k \bigl(\frac{1}{2}-d\bigr)} u_{n}(k,x).
\end{align*}
Then, since
\begin{align*}
\iu \langle \tilde{u}_{n}(k,\cdot)|\partial_k \tilde{u}_{n}(k,\cdot)\rangle = \iu \langle u_{n}(k,\cdot)|\partial_k u_{n}(k,\cdot)\rangle -\tfrac{1}{2}+d,
\end{align*}
the Zak phase associated to the new unit cell is given by
\begin{align}\label{eq:Zd}
\tilde{Z}_n(d)=Z_n-\pi(1-2d),
\end{align}
where $Z_n=\tilde{Z}_n(\tfrac{1}{2})$.

\subsubsection{Numerical calculation}
In order to compute numerically the Zak phase, we discretize the Brillouin zone $[-\pi,\pi)$ by considering the $M$ points 
\begin{align*}
k_i=-\pi+i\frac{2\pi }{M}
\end{align*}
for $i\in\{0,\dots ,M\}$. The Zak phase is then given by~\cite{Res94, Guo16}
\begin{align*}
Z=-\lim_{M\to +\infty} \Im \log  \prod_{i=0}^{M-1} S_i,&& S_i=\langle u(k_i,\cdot)|u(k_{i+1},\cdot )\rangle.
\end{align*}
where we dropped the dependence on the quantum number $n$ to declutter the notations. For the case discussed in the main text, see Eqs.~\eqref{eq:eigenspinor} and~\eqref{eq:Zakeig},  the inner product $S_i$ has four contributions, which can be computed explicitly. Setting  $\xi_i^\pm=\xi^\pm|_{k=k_i}$ and $q_i=q(\epsilon(k_i))$  we have that
\begin{align*}
    S_i=S_i^{(1)}+S_i^{(2)}+S_i^{(3)}+S_i^{(4)},
\end{align*}
where 
\begin{align}\label{eq:Si} 
    S_i^{(j)}=A_i^{(j)}\biggl( \int_{-\frac{1}{2}}^{0} \e^{\iu \alpha_i^{(j)} x}\,\dd x+\int_{0}^{\frac{1}{2}} \e^{\iu \alpha_i^{(j)} (x-1)}\,\dd x\biggr)
    =A_i^{(j)}\e^{-\iu \frac{\alpha_i^{(j)}}{2}}\sinc\biggl(\frac{\alpha_i^{(j)}}{2}\biggr)
\end{align}
and 
\begin{align*}
    A_i^{(1)}=(\xi_{i}^+)^\dagger \xi_{i+1}^+, &&\alpha_i^{(1)}&=q_{i+1}-\overline{q}_{i}+\tfrac{2\pi}{M},\\
    A_i^{(2)}=(\xi_{i}^-)^\dagger \xi_{i+1}^+, &&\alpha_i^{(2)}&=q_{i+1}+\overline{q}_{i}+\tfrac{2\pi}{M},\\
    A_i^{(3)}=(\xi_{i}^+)^\dagger \xi_{i+1}^-, &&\alpha_i^{(3)}&=-q_{i+1}-\overline{q}_{i}+\tfrac{2\pi}{M},\\
    A_i^{(4)}=(\xi_{i}^-)^\dagger \xi_{i+1}^-, &&\alpha_i^{(4)}&=-q_{i+1}+\overline{q}_{i}+\tfrac{2\pi}{M}.
\end{align*}

\end{document}